\documentclass[12pt,a4]{article}
\usepackage{relsize}
\usepackage{array}
\usepackage{a4wide}
\usepackage[hidelinks]{hyperref}
\usepackage{latexsym}
\usepackage{cite}

\usepackage{caption}
\usepackage{mathrsfs}
\usepackage{tikz}
\usetikzlibrary{arrows.meta,calc,decorations.markings,math,arrows.meta,shapes.misc, decorations.pathreplacing}
\usepackage{amssymb}
\usepackage{amsmath}
\usepackage{graphicx}
\usepackage{enumerate}
\usepackage{todonotes}
\usepackage{a4wide}
\usepackage{tensor}
\usepackage{comment}
 
\usepackage{filecontents}

\newcommand{\tr}{\mathrm{tr}}

\newcommand{\be}{\begin{equation}\label}
\newcommand{\ee}{\end{equation}}
\newcommand{\bea}{\begin{eqnarray}\label}
\newcommand{\eea}{\end{eqnarray}}

\makeatletter
\newcommand*{\textoverline}[1]{$\overline{\hbox{#1}}\m@th$}
\newcommand*\bigcdot{\mathpalette\bigcdot@{.65}}
\newcommand*\bigcdot@[2]{\mathbin{\vcenter{\hbox{\scalebox{#2}{$\m@th#1\bullet$}}}}}
\makeatother

\numberwithin{equation}{section} 

\begin{document}

\newcommand{\PR}[1]{{\color{red} [PR: #1]}}
\newcommand{\RM}[1]{{\color{red} [RM: #1]}}
\newcommand{\NL}[1]{{\color{red} [NL: #1]}}
\newcommand{\AL}[1]{{\color{red} [AL: #1]}}

\newcommand{\nn}{\nonumber}

 \centerline{\LARGE \bf {\sc  
 Instanton Worldlines In} } 
 \vspace{12pt}
 \centerline{\LARGE \bf {\sc Five-Dimensional $\Omega$-Deformed
  Gauge Theory} } 

  \vspace{1cm}
  \centerline{
   {\large {\bf  {\sc N.~Lambert,${}^{\,a}$}}\footnote{E-mail address: \href{neil.lambert@kcl.ac.uk}{\tt neil.lambert@kcl.ac.uk}}     \,{\sc A.~Lipstein,$^{\,b}$}\footnote{E-mail address: \href{mailto:arthur.lipstein@durham.ac.uk}{\tt arthur.lipstein@durham.ac.uk}}\, {\sc R.~Mouland${}^{\,a}$}\footnote{E-mail address: \href{rishi.mouland@kcl.ac.uk}{\tt rishi.mouland@kcl.ac.uk}}   {\sc    and P.~Richmond${}^{\,a}$}}\footnote{E-mail address: \href{mailto:paul.richmond@kcl.ac.uk}{\tt paul.richmond@kcl.ac.uk}}  }  
     
\vspace{1cm}
\centerline{${}^a${\it Department of Mathematics}}
\centerline{{\it King's College London }} 
\centerline{{\it The Strand, WC2R 2LS, UK}} 
  
\vspace{1cm}
\centerline{${}^b${\it Department of Mathematical Sciences}}
\centerline{{\it Durham University}} 
\centerline{{\it  Durham, DH1 3LE, UK}}

\vspace{1.0truecm}

 
\thispagestyle{empty}

\centerline{\sc Abstract}
\noindent We discuss  the Bosonic sector 
of a class of supersymmetric non-Lorentzian five-dimensional gauge field theories with an $SU(1,3)$ conformal symmetry. These actions have a Lagrange multiplier which imposes a novel $\Omega$-deformed anti-self-dual gauge field constraint. Using a generalised 't Hooft ansatz we find the constraint equation linearizes allowing us to construct a wide class of explicit solutions. These include finite action configurations that describe worldlines of anti-instantons which can be created and annihilated. We also describe the dynamics on the constraint surface. 
\newpage
\tableofcontents
\section{Introduction}

One of the more interesting predictions to arise from String theory and M-theory is the existence of a rich spectrum of superconformal gauge field theories above four dimensions. The possibility of such theories first arose in the classification of Nahm \cite{Nahm:1977tg}. Much later physical realisations of these theories were constructed from strongly coupled branes in String and M-theory, starting with \cite{Witten:1995zh,Strominger:1995ac,Seiberg:1996qx}. Thus they claim to have a fundamental importance in M-theory. They also present an important challenge to our general understanding of quantum field theory and its myriad of applications.

These theories are not expected to have a Lagrangian description, at least not in any standard sense. In recent work we have obtained a class of five-dimensional Lagrangian gauge field theories without Lorentz invariance but with up to 3/4 of the maximal supersymmetry \cite{Lambert:2019jwi,Lambert:2020jjm}. They were obtained by considering null reductions of six-dimensional superconformal field theories on an $\Omega$-deformed Minkowski space, which in turn arises as a conformal compactification of six-dimensional Minkowski space. The hope is that they can be used to formulate a Lagrangian description of at least a part of the parent six-dimensional theory. For this reverse construction to work we must interpret the instanton number of the gauge fields on the spatial sections as the missing null Kaluza-Klein momentum. This is in line with previous DLCQ constructions \cite{Aharony:1997th,Aharony:1997pm} but with better control \cite{Lambert:2020zdc}. 

However these theories appear  to be interesting in their own right. Indeed the class of such five-dimensional non-Lorentzian theories is bigger than those that are obtained from reduction of a parent six-dimensional theory with $(2,0)$ or $(1,0)$ superconformal symmetry. For example these five-dimensional actions place no restrictions on the gauge group and matter representations whereas these are tightly constrained in the six-dimensional superconformal field theories.  Furthermore, in addition to their supersymmetry, these Lagrangians admit an $SU(1,3)$ spacetime symmetry \cite{Lambert:2019fne}. This plays a similar role to the $SO(2,5)$ conformal symmetry group of a Lorentzian five-dimensional theory. In particular the $SU(1,3)$ symmetry contains the following generators:
\begin{itemize}
  \item A non-Abelian subgroup of five translations, whose finite action nonetheless can map between any two points in the five-dimensional spacetime.	\vspace{-2mm}
  \item A subgroup of four rotations in the four $x^i$ directions. \vspace{-2mm}
  \item A Lifshitz scaling symmetry, under which the fifth coordinate $x^-$ scales twice as quickly as the $x^i$. \vspace{-2mm}
  \item Five `special' transformations, which in the representation theory play a role analogous to the special conformal transformations of the conformal group.
\end{itemize} 
In a recent paper we have shown that this symmetry places non-trivial constraints on the correlation functions \cite{Lambert:2020zdc}. 

In this paper we will initiate an analysis of the dynamical aspects of these Lagrangians. In particular, a first step in this direction is to understand the anti-self-duality constraint that arises from integrating out a Lagrange multiplier field from the action. Using a generalised 't Hooft ansatz we will find a rich class of solutions to the anti-self-duality constraint and describe how the dynamics of the theory reduces to a theory of interacting instanton-solitonic particle worldlines. Indeed we are able to use the 't Hooft ansatz to solve for many of the dynamical Bosonic fields in the theory.

The rest of this paper is organised as follows. In Section 2 we review the action and $\Omega$-deformed anti-self-dual gauge field constraint. In Section 3 we show how to solve this constraint using a generalised 't Hooft ansatz. The result is an infinite class of solutions where the size and position moduli of the undeformed case are allowed to evolve along the worldline of the anti-instanton, resulting in a modified gauge field which can be viewed as incorporating the backreaction. We argue that these solutions include worldlines of anti-instantons that can be created and annihilated and in Section 4 we provide a discussion of topological aspects of the gauge field. In Section 5 we discuss the solutions of the scalar fields in the presence of these anti-instanton worldline solutions. In Section 6 we discuss the dynamics of the system on the constraint surface. Section 7 contains our conclusions and a discussion.

\section{The Action and Generalised Anti-Self-Dual Constraint}

The Bosonic part of the actions we wish to study can be written as \cite{Lambert:2019jwi,Lambert:2020jjm}
 \begin{align}
\begin{split}\label{action}
S = \frac{1}{{g^2_{\text{YM}}}} \int dx^{-} d^4 x \Big \{ \frac{1}{2} \tr \big( F_{i-} F_{i-}  \big ) + \frac{1}{2} \tr \big (\mathcal{F}_{ij} G^+_{ij} \big ) -\frac{1}{2} \tr \big (\hat D_{i} \phi\hat D_{i} \phi \big ) -   \hat D_i {X}_{\alpha}{}^{m}\hat D_i X^{\alpha}{}_{m}  \Big\}\ ,
\end{split}
\end{align}
where $i,j=1,2,3,4$. Here  ${X}^{\alpha}{}_{m} = ({X}_{\alpha}{}^{m})^\dag$ are the scalar fields from   six-dimensional hyper-multiplets labelled by $m$, which can sit in any representation of the gauge group. On the other hand 
$A_-,A_i$ are components of a five-dimensional gauge field with field strength $F_{-i},F_{ij}$ and  $G^+_{ij}=\tfrac12\varepsilon_{ijkl}G^+_{kl}$ whereas $\phi$ is a scalar. Altogether $A_-,A_i, G^+_{ij}$ and $\phi$   arise as the Bosonic fields in  a null reduction of a six-dimensional tensor multiplet  and take values in the adjoint of the gauge group. Lastly we define
\begin{align}
\hat D_i &= D_i - \tfrac12 \Omega_{ij}x^jD_- \ , \nn\\
  \mathcal{F}_{ij}  &= F_{ij}	 - \frac12\Omega_{ik}x^kF_{-j}+\frac12\Omega_{jk}x^kF_{-i}\ ,
\end{align}
where $D_-, D_i$ are adjoint gauge covariant derivatives for the gauge field $A_-, A_i$, {\it i.e.}\ $D_-=\partial_- - i [A_-,\,\cdot\,]$ and $D_i=\partial_i - i [A_i,\,\cdot\,]$.  Furthermore, $\Omega_{ij}$ is an anti-self-dual 2-form such that $\Omega_{ik}\Omega_{jk}=R^{-2}\delta_{ij}$ and $R$ is a constant with dimensions of length. 

This action is invariant under an $SU(1,3)$ conformal symmetry \cite{Lambert:2019fne} and, when fermions are included, $4$ supersymmetries and $8$ superconformal symmetries. In special cases, corresponding to six-dimensional $(2,0)$ theory, this is enhanced to  $8$ supersymmetries and $16$ superconformal symmetries \cite{Lambert:2019jwi,Lambert:2020jjm}.

The self-dual field $G^+_{ij}$ acts as a Lagrange multiplier imposing the constraint
\begin{align}
  \mathcal{F}_{ij} =- \star \mathcal{F}_{ij}\ ,  \label{eq: instanton eqn bosonic}
  \end{align}
  where $\star$ is the Hodge dual on ${\mathbb R}^4$.
Our task here is to analyse solutions to this constraint and explore the dynamics arising from the action $S$. 

Let us introduce
\begin{align}
\hat A_i &= A_i - \frac12 \Omega_{ij}x^j A_- \ , \nonumber\\
\hat \partial_i & = \partial_i - \frac12\Omega_{ij}x^j\partial_-	\ .
\end{align}
Note that $\hat \partial_i$ has torsion:
\begin{align}
[\hat\partial_i,\hat\partial_j]=\Omega_{ij}\partial_-\ .
\end{align} 
It also follows from this that 
 if $\hat\partial_i f=0$ then $\partial_-f=0$ and hence $f$ is constant. 

With these definitions 
we can write the $\Omega$-deformed field strength as
\begin{align}
{\cal F}_{ij}	&= \hat{ F}_{ij} -  \Omega_{ij}A_- \ , \nonumber\\
\hat{ F}_{ij}&=\hat\partial_i \hat A_j -\hat\partial_j\hat A_i - i[\hat A_i,\hat A_j]\ .
\end{align}
Thus the anti-self-duality condition becomes a constraint on $\hat A_i$: $\hat{ F}_{ij} =-\star \hat{ F}_{ij} $.  

Note that for $x^-$-independent solutions   $\hat A_i$ can be viewed as an ordinary anti-self-dual gauge field and  $\hat{F}_{ij}$ its field strength. Thus every $x^-$-independent solution to the familiar anti-self-dual gauge field condition, given by the ADHM construction in terms of moduli, is also a solution to the  $\Omega$-deformed anti-self-dual gauge condition. 

However,   when there is a non-trivial dependence on $x^-$, the $\Omega$-deformed anti-self-duality condition is more restrictive since it constrains the dependence on $x^-$ whereas in the undeformed theory there are no constraints on the $x^-$-dependence of the moduli. Nevertheless we are able to construct an infinite dimensional space of $x^-$-dependent solutions to the   $\Omega$-deformed anti-self-dual gauge condition using a generalisation of the 't Hooft Ansatz. These carry a non-trivial dependence on $x^-$ along with a modified dependence on $x^i$ reminiscent of a backreaction effect.

\section{Generalised 't Hooft Ansatz}

It would be interesting to obtain an ADHM-like construction for the general solution to  $\hat{ F}+\star\hat{ F}=0$. In lieu of this we restrict attention to an $SU(2)$ gauge group and make the 't Hooft-like ansatz
\begin{align}
\hat A_i = \eta^a_{ij}\partial_j B\sigma^a + \Omega_{ij}\eta^a_{jk} C_k\sigma^a\ .
 \end{align} 
 Here $\eta^a_{ij}$, $a=1,2,3$ are a basis for self-dual 2-forms and $\sigma^a$ are the Pauli matrices. For a nice review and some useful formulae see \cite{Vandoren:2008xg}.
 
With this ansatz we find the conditions for $\hat{ F}_{ij}+\star\hat{ F}_{ij}=0$ are
\begin{align}
  -2\partial_i\partial_i B + 4 \partial_i B \partial_i B + \Omega_{ij} x^j \partial_-\partial_i B + R^{-2} x^i \partial_- C_i + 8 \Omega_{ij} \partial_i B\, C_j + 4R^{-2} C_i C_i - 2\Omega_{ij} \partial_i C_j &= 0 \ , \nonumber\\
\eta^c_{ki}\Omega_{kj}\left( -x^j \partial_-\partial_i B + \Omega_{li} x^j \partial_- C_l - 2\partial_i C_j \right) &= 0\ .
\end{align} 
We can solve the second equation by simply taking
\begin{align}
C_i = -\frac 12 x^i \partial_- B	\ .
\end{align}
In this case we find that $\hat{ F}=-\star\hat{ F}$ only requires the first equation to vanish. This becomes
\begin{align}
  -2\left( \partial_i \partial_i B - 2\partial_i B \partial_i B \right) + 2\Omega_{ij} x^j \left( \partial_i \partial_- B - 2 \partial_- B \partial_i B  \right) - \frac{|\vec x|^2 }{2R^2} \left( \partial_- \partial_- B - 2\partial_- B \partial_- B \right)=0\ ,
\end{align}
with $|\vec x|^2=x^i x^i$.
Hence, introducing $B=-\frac{1}{2}\log\Phi$, we have
\begin{align}
  \partial_i \partial_i\Phi - \Omega_{ij} x^j \partial_-\partial_i \Phi + \frac{1}{4}R^{-2} |\vec x|^2 \partial_-\partial_- \Phi = 0 \quad\Longleftrightarrow\quad \hat\partial_i\hat\partial_i \Phi= 0\ ,
\label{eq: linearised eqn}
\end{align}
and
\begin{align}
\hat A_i = 
-\frac12\eta^a_{ij}\sigma^a\hat\partial_j \ln \Phi \ .
\label{eq: 't Hooft form for A}
 \end{align} 
Note that it follows from this that $\hat \partial_i\hat A_i=\hat D_i\hat A_i=0$ as $\Omega_{ij}\eta^a_{ij}=0$. One can also compute 
\begin{align}
	\hat{ F}_{ij} = \frac12\Phi \hat \partial_{i}\hat \partial_{k} \Phi^{-1}\eta^a_{jk}\sigma^a-\frac12\Phi \hat \partial_{j}\hat \partial_{k} \Phi^{-1}\eta^a_{ik}\sigma^a+ \frac12\Phi^2\hat\partial_k\Phi^{-1} \hat\partial_k\Phi^{-1} \eta^a_{ij}\sigma^a\ .
\end{align}

Thus we find that using a 't Hooft ansatz the self-duality condition reduces to a linear second order differential equation for $\Phi$. This suggests that a more general ADHM construction could also be obtained.

\subsection{Spherically Symmetric Solutions}

Let us start by looking for spherically symmetric solutions where $\Phi=\Phi\left( x^-, |\vec x|^2 \right)$.  Then we find
\begin{align}
  4|\vec x|^2\frac{\partial^2\Phi}{\partial\left( |\vec x|^2 \right)^2} + 8\frac{\partial\Phi}{\partial|\vec x|^2} + \frac{ |\vec x|^2}{4R^2} \frac{\partial^2\Phi}{\partial({x^-})^2}  = 0\ .
\end{align}
This can be written in a nicer form by defining the complex variable
\begin{align}
 z=x^- + \frac{i}{4R} |\vec x|^2. 
\end{align}
Then we have
\begin{align}
  \left( z-\bar{z} \right)\partial \bar{\partial} \Phi - \left( \partial - \bar{\partial} \right) \Phi = \partial \bar{\partial}\big((z-\bar{z})\Phi\big) = 0\ .
\end{align}
Hence, the \textit{general} spherically symmetric solution of $\Phi$ is given by
\begin{align}
  \Phi = \frac{1}{z-\bar{z}} \Big(\, \varphi_+\left( z \right) + \varphi_-\left( \bar{z} \right)\Big)\ .
\end{align}
However Hermiticity of $\hat A_i$ requires that   $\Phi$ is real and hence imposes  $\varphi_-= -\bar{\varphi}_+$ and so 
\begin{align}
  \Phi = \frac{1}{z-\bar{z}} \Big(\, \varphi_+\left( z \right) - \bar \varphi_+\left( \bar{z} \right)\Big)\ .
\end{align}

Furthermore to avoid serious singularities in $\hat A_i$  we also require that $\Phi >0$. Since the imaginary part of $z$ is positive definite this requires that ${\rm Im}(\varphi_+)>0 $ on the upper half-plane. According to \cite{Aronszajn1956} the general solution can be written as  \begin{align}
\varphi_+(z) &=\alpha +   \gamma^2(z-{\rm Re}(z_0)) 	+\int^\infty_{-\infty}\left(\frac{1}{\tau-z} - \frac{\tau -{\rm Re}(z_0)}{|\tau - z_0|^2} \right)\mu(\tau)d\tau\ ,
\end{align}
where $z_0$ lies in the complex upper half-plane but $\alpha,\beta,\gamma$ and $\tau$ are real and $\mu(\tau)\ge 0$ is arbitrary so long as the integral exists. Note that, assuming $\gamma\ne 0$, we can rescale $\Phi$ to set $\gamma=1$ without altering the gauge field $\hat A_i$. Thus we see that $\Phi$ takes the form
\begin{align}
 \Phi & =  1  + \int^\infty_{-\infty} \frac{\mu(\tau)}{|\tau-z|^2}  d\tau\ .
 \label{eq: Phi spherically symmetric solution}
\end{align}
In particular, $\Phi$ is regular except for at the line of points at $x^i=0$ when $\mu(x^-)>0$. As we will see in more detail in Section \ref{sec: gauge field topology}, these lines can be seen as the worldlines of single anti-instantons ({\it i.e.}\ instantons with degree $-1$), with points at which $\mu(x^-)$ transitions between $\mu(x^-)=0$ and $\mu(x^-)>0$ interpreted as their creation or annihilation. Such configurations can be understood in the quantum theory as arising from particular local boundary conditions in the path integral, or equivalently in terms of instanton operators \cite{Lambert:2014jna,Tachikawa:2015mha,Bergman:2016avc,US}.

Finally, it is worth briefly noting that the inclusion of $\Omega_{ij}\neq 0$ introduces a preferred duality relation, which in particular breaks the straightforward symmetry between solutions of $\hat{F}_{ij}=+\star \hat{F}_{ij}$ and $\hat{F}_{ij}=-\star \hat{F}_{ij}$ that is present when $\Omega_{ij}=0$. To see this, let us firstly briefly review the case $\Omega_{ij}=0$, where $\hat{F}_{ij}=F_{ij}$ is the usual field strength of the gauge field $A_i$. Solutions of $F_{ij}=-\star F_{ij}$ with generic instanton number $k<0$ are found in singular gauge as\footnote{One could alternatively try this ansatz with the anti-self-dual 't Hooft matrices $\bar{\eta}^a_{ij}$, but would only find the $k=-1$ instanton in regular gauge.} $A_i=-\frac12\eta^a_{ij}\sigma^a\partial_j \ln \Phi$ for harmonic $\Phi$. To find solutions to the opposite equation $F_{ij}=\star F_{ij}$ with $k>0$, one simply swaps $\eta^a_{ij} \to \bar{\eta}^a_{ij}$.

Let us now go back to $\Omega_{ij}\neq 0$, and take $\Omega_{ij}$ anti-self-dual as we do throughout this paper. In solving $\hat{F}_{ij}=\pm\star \hat{F}_{ij}$, we may in principle consider two different ans\"{a}tze: $\hat{A}_i=-\frac12\eta^a_{ij}\sigma^a\hat\partial_j \ln \Phi$, or $\hat{A}_i=-\frac12\bar{\eta}^a_{ij}\sigma^a\hat\partial_j \ln \Phi$. As we have seen, it is the former ansatz involving $\eta^a_{ij}$ that proves fruitful in solving $\hat{F}_{ij}=-\star \hat{F}_{ij}$. One might then hope that the latter ansatz involving $\bar{\eta}^a_{ij}$ will be similar useful in solving $\hat{F}_{ij}=+\star \hat{F}_{ij}$. This is however not the case. In particular, considering the following two parameterisations,
\begin{align}
\Phi=\frac{g_{A}(z,\bar{z})}{z-\bar{z}}=\left(z-\bar{z}+g_{B}(z,\bar{z})\right)^{-1},
\end{align}
we find the following constraints depending on which equation we are trying to solve, and which ansatz we are using\footnote{Note, if we had instead chosen $\Omega_{ij}$ self-dual, we need simply to swap the rows and columns of this table, so that each entry is exchanged with its diagonal opposite.}:
\[
\begin{array}{l|cc}
 & \eta & \bar{\eta}\\
 \hline\hat{F} =-\star\hat{F}  & \partial\bar{\partial}g_{A}=0 & \partial\bar{\partial}g_{B}=\partial^{2}g_{B}=\bar{\partial}^{2}g_{B}=0\\
\hat{F} =\star\hat{F}  & \partial\bar{\partial}g_{B}=\partial^{2}g_{B}=\bar{\partial}^{2}g_{B}=0 & \partial\bar{\partial}g_{A}=\partial_{-}g_{A}=0.
\end{array}
\]
The important take away is that the top left and bottom right entries are qualitatively different, and in particular the $\bar{\eta}$ ansatz gives only static solutions to $\hat{F}_{ij}=\star\hat{F}_{ij}$. In this way, we see that the anti-self-duality of $\Omega_{ij}$ breaks the symmetry between $\hat{F}_{ij}=\star\hat{F}_{ij}$ and $\hat{F}_{ij}=-\star\hat{F}_{ij}$.

\subsection{Simple Examples}

Before we continue let us first look at some simple forms for $\Phi$. If $\mu = \rho^2/4\pi R$ is constant, then we find from \eqref{eq: Phi spherically symmetric solution}
\begin{align}\label{S'tH}
\Phi = 1 + \frac{\rho^2}{|\vec x|^2}	\ .
\end{align}
This gives back the usual static instanton located at $x^i=0$ and with size $\rho$. A more interesting example is
\begin{align}
\mu(\tau) = \frac{\rho^2_0l^2/4\pi R}{(\tau-\tau_0)^2+l^2}	\ ,
\end{align}
which leads to
\begin{align}\label{ex2}
	\Phi = 1 + \frac{\rho^2_0}{|\vec x|^2}\frac{l(l+\tfrac{1}{4R}|\vec x|^2)}{(x^--\tau_0)^2 + (l+\tfrac{1}{4R}|\vec x|^2)^2}\ .
\end{align}
Here the small $|\vec x|$ behaviour is unchanged except that the instanton size grows and then decays in $x^-$. Note that  $\Phi \sim 1/|\vec x|^4$ as $|\vec x|\to\infty$. Taking the limit $l\to\infty$ leads to the static instanton.  On the other hand taking the limit $l\to 0$ with $\rho^2=\rho^2_0 l/4R $ fixed gives
\begin{align}\label{ex3}
	\Phi = 1 + \frac{\rho^2} {(x^--\tau_0)^2 + \tfrac{1}{16R^2}|\vec x|^4}\ .
\end{align}
Which corresponds to  $\mu(\tau) = \rho^2\delta(\tau-\tau_0)$ and does not lead to an instanton  as $\Phi$ is smooth as a function of $x^i$ except at $x^-=\tau_0$ where it produces a singular gauge field. 

We can also consider a simple oscillating  anti-instanton by taking $\mu(\tau) = A-B\cos(\tau)$ with $A\ge B>0$:
\begin{align}
	\Phi = 1 +  {4\pi R }\frac{A-Be^{-{|\vec x|^2/4R}}\cos(x^-)}{|\vec x|^2}.
\end{align}
Note that taking $A=B$ the small $|\vec x|$ limit gives
\begin{align}\label{wavey}
	\Phi = 1 +  {4\pi A R }\frac{1- \cos(x^-)}{|\vec x|^2}+\ldots\ ,
\end{align}
corresponding to instantons that shrink to zero size and then grow again.

Another $x^-$-dependent example is simply  a step function
\begin{align}
\mu(\tau) = \begin{cases}
0 & \tau <\tau_1 \\
\rho^2/4\pi R &  \tau_1\le \tau\le \tau_2\\
0 &  \tau_1 <\tau 	\\
 \end{cases}\ ,
\end{align}
 so that
\begin{align}
\Phi  =&\  1 +\frac{\rho^2}{ |\vec x|^2}\frac{1}{2\pi i}\left[\ln \left(\frac{\tau_2 - z}{\tau_2 - \bar z}\right)-\ln \left(\frac{\tau_1 - z}{\tau_1 - \bar z}\right)\right]	 \ .\end{align} 
Note that in doing so we must choose a branch of the logarithm such that for $z$ in the upper half-plane, $\ln (z/\bar{z})=2i\arg(z)$, with $0\le \arg(z) \le \pi$. The logarithms are bounded and regular everywhere except for $(x^-,x^i)=(\tau_1,0)$ and  $(x^-,x^i)=(\tau_2,0)$. As such  for $x^i\to0$ we find 
\begin{align}
	\ln \left(\frac{\tau-z}{\tau - \bar z}\right)\to \begin{cases}
2\pi i   & x^-<\tau  \\ 
0 &x^->\tau	
 \end{cases}\ ,
\end{align}
but note that if $x^->\tau_2$ then  $x^->\tau_1$ and if $x^-<\tau_1$ then  $x^-<\tau_2$ hence, as $|\vec x|^2\to0$,
\begin{align}
	\ln \left(\frac{\tau_2 - z}{\tau_2 - \bar z}\right)-\ln \left(\frac{\tau_1 - z}{\tau_1 - \bar z}\right)\to \begin{cases}
0  & x^->\tau_{2}  \\ 
2\pi i  &\tau_1<x^-<\tau_{2}\\
	0  & x^-<\tau_{1}
 \end{cases}\ .
\end{align}
Thus we   create an instanton centred at the origin at $x^-=\tau_1$ and destroy it at $x^-=\tau_2$.  
Taking $\tau_1\to-\infty$ and $\tau_2\to\infty$ we recover the static solution.  
However this solution has infinite action (at least when $A_-=0$) arising from the delta-function in $d\mu/d\tau$. 

A smoother, continuous, example is
  ($\tau_4>\tau_3>\tau_2>\tau_1$)
 \begin{align}
  \mu(\tau)=\left\{\begin{aligned}
  &\,\, 0 &\tau < \tau_1		\nn\\
  &\,\,\frac{\rho^2}{4\pi R }\left( \frac{\tau-\tau_1 }{\tau_2-\tau_1}\right)\qquad	&\tau_1 \le  \tau \le  \tau_2\nn\\
  &\,\,\frac{\rho^2} {4 \pi R}	\qquad	&\tau_2<\tau<\tau_3	\nn\\
    &\,\,  \frac{\rho^2}{  4\pi R }\left( \frac{\tau_4-\tau }{\tau_4-\tau_3}\right)	\qquad	&\tau_3 \le \tau \le \tau_4	\nn\\
  &\,\,0 &\tau >    \tau_4		
\end{aligned}\right.\ .
\end{align}
The resulting $\Phi$ takes the rather ugly form
\begin{align}
\Phi  = 1 &+\frac{\rho^2}{|\vec x|^2} \frac{ 1}{2\pi i} \frac{1}{\tau_2-\tau_1}\left[ z\ln \left(\frac{\tau_2-z}{\tau_1  -  z}\right) -\bar z\ln\left(\frac{\tau_2-\bar z}{\tau_1 - \bar z}\right) \right] \nonumber\\
&-\frac{\rho^2}{|\vec x|^2} \frac{ 1}{2\pi i} \frac{\tau_1}{\tau_2-\tau_1}\left[ \ln\left(\frac{\tau_2 - z}{\tau_2- \bar  z}\right) - \ln\left(\frac{\tau_1 - z}{\tau_1 -\bar  z}\right)\right]\nonumber\\
&+\frac{\rho^2}{|\vec x|^2} \frac{ 1}{2\pi i}\left[   \ln\left(\frac{\tau_3- z}{\tau_3 -  \bar z}\right)-\ln\left(\frac{\tau_2- z}{\tau_2 -  \bar z}\right) \right]\nonumber\\
&+\frac{\rho^2}{|\vec x|^2} \frac{ 1}{2\pi i}\frac{\tau_4}{\tau_4-\tau_3}\left[   \ln\left(\frac{\tau_4 - z}{\tau_4- \bar  z}\right) - \ln\left(\frac{\tau_3 - z}{\tau_3 -\bar  z}\right)\right]\nonumber\\
&-\frac{\rho^2}{|\vec x|^2} \frac{ 1}{2\pi i}\frac{1}{\tau_4-\tau_3}\left[ z\ln \left(\frac{\tau_4-z}{\tau_3  -  z}\right) -  \bar z\ln \left(\frac{\tau_4-\bar z}{\tau_3  - \bar  z}\right)\right]	\ .
\end{align}
However one can see that,  since $\bar z = z - \frac{i}{2R}|\vec x|^2$,  the first and last lines are finite as $|\vec x|\to 0$, whereas the middle lines behave similarly to the previous case. Thus we find a smooth finite action (at least for $A_-=0$) solution to the constraint  that represents the creation and then annihilation of an instanton at $\vec x=0$.

\subsection{Allowing General Worldlines}

We can significantly generalise the spherically symmetric solution (\ref{eq: Phi spherically symmetric solution}). Since the equation for $\Phi$ is linear we can obtain new solutions by summing over existing solutions. However any sum over spherically symmetric solutions remains spherically symmetric and hence just changes  the form of the function $\mu$. To find more solutions we can leverage the $SU(1,3)$ spacetime symmetry enjoyed by the theory. 
The behaviour of the action under these symmetries is subtle in the presence of instantons. However 
it turns out that  the constraint equation $\mathcal{F}_{ij} + \star \mathcal{F}_{ij}=0$ is manifestly $SU(1,3)$ invariant \cite{US}. 

In particular we can consider  translations. These take the form
\begin{align}
	x'^- = x^- - y^- + \frac12 \Omega_{ij}x^i y^j \ , \qquad x'^i = x^i - y^i\ .
	\label{eq: translated coordinates}
\end{align}  If we let
\begin{align}
\Phi'(x^-,x^i) 
	= \Phi(x'^-,x'^i)\ ,
\end{align}
then we find that
\begin{align}
\hat  \partial_i \hat \partial_i\Phi'&=  \partial_i \partial_i\Phi' - \Omega_{ij} x^j \partial_-\partial_i \Phi' + \frac{1}{4R^2}  |\vec x|^2 \partial_-^2 \Phi'\nonumber\\ & =   
   \partial'_i \partial'_i\Phi - \Omega_{ij} x'^j\partial'_-\partial'_i \Phi + \frac{1}{4R^2} |\vec x'|^2 {\partial'}_-^2\Phi\nonumber \\
   &= \hat  \partial'_i \hat \partial'_i\Phi'\ .
  \end{align}
Hence if $\Phi$ satisfies (\ref{eq: linearised eqn}) then so does $\Phi'$. Thus the translations (and also the Lifshitz scaling) preserve the 't Hooft form (\ref{eq: 't Hooft form for A}). Let us use these translations to derive a significant generalisation of (\ref{eq: Phi spherically symmetric solution}).

We note that the solution (\ref{eq: Phi spherically symmetric solution}) is a continuous linear sum over solutions of the form $(z-\tau)^{-1}(\bar{z}-\tau)^{-1}$, for any $\tau\in\mathbb{R}$. Indeed, by taking $\mu(\tau)=\rho^2\delta(\tau)$ we have simply
\begin{align}
\Phi = 1 + \frac{\rho^2}{z\bar z}	\ .
\end{align}
Using (\ref{eq: translated coordinates}), we can then translate this solution to find a new solution
\begin{align}
\Phi = 1 + \frac{\rho^2}{ z(x,y)\bar{z}(x,y)}	\ ,
\end{align}where here for any pair of points $x=(x^-, x^i),y=(y^-, y^i)$ in our spacetime, we define
\begin{align}
  z(x,y)=x^-- y^- + \frac{1}{2}\Omega_{ij} x^i y^j + \frac{i}{4R}|\vec x - \vec y|^2\ .
\end{align} 
In particular $z=z(x,0)$, and note that $z(x,y)=0$ only if $x=y$.
The real and imaginary parts of $z(x,y)$ constitute the unique translationally-invariant quantities. Indeed, $z(x,y)$ defines a very useful covariant distance, which appears for instance heavily in the form of correlation functions \cite{Lambert:2020zdc}.

Finally, in order that $\Phi$ continues to describe a particle-like configuration, we integrate over a one-parameter family of the translated solutions, each centred at some point $y(\tau)=(y^-(\tau), y^i(\tau))$. The result is the solution
\begin{align}
  \Phi = 1+ \int d\tau \frac{\mu(\tau)}{z(x,y(\tau))\bar{z}(x,y(\tau))}\ ,
  \label{eq: Phi general single particle}
\end{align}
where $\mu(\tau) \ge 0$.

Let us now interpret this solution. We once again find this solution describes a particle-like ({\it i.e.}\ co-dimension four) object. This is seen by noting that $\Phi$ and hence $\hat{A}_i$ is singular precisely at any point $x$ such that there exists some $\tau$ with $x=y(\tau)$ and $\mu(\tau)>0$. We see that the spacetime curve $y(\tau)=(y^-(\tau), y^i(\tau))$ is precisely the worldline of this particle, with $\tau$ providing a local parameterisation along it.
We  recover the spherically symmetric solution (\ref{eq: Phi spherically symmetric solution}) by considering the case $y^i(\tau)=0$ and  $y^-(\tau)=\tau$. 

Note that we are free to sum up $N$ disjoint  particle-like configurations for $\Phi$. Thus a  yet more general solution is then given by
\begin{align}
  \Phi = 1+ \sum_{A=1}^N\int d\tau_A \frac{\mu_A(\tau)}{z(x,y_A(\tau_A))\bar{z}(x,y_A(\tau))}\ ,
  \label{eq: Phi general solution}
\end{align}
for any $N$. Such a solution then describes not one but a  swarm   of $N$ instanton particles. Note that such solutions can nonetheless be brought back to the form (\ref{eq: Phi general single particle}) by connecting each worldline end-to-end with new segments along which $\mu(\tau)=0$. As such, the form (\ref{eq: Phi general solution}) is an equivalent rather than generalised form for $\Phi$, which is nonetheless a useful representation of the solution.

Let us briefly comment on the gauge group embedding. If we take a solution
to the anti-self-duality constraint $\hat{F}_{ij}=-\star \hat{F}_{ij}$ and do the following
\begin{align}
\hat{A}_{i}\rightarrow U^{-1}\hat{A}_{i}U,
\end{align}
where $U(x^{-})\in SU(2)$, then the field strength becomes
\begin{align}
\hat{F}_{ij}\rightarrow U^{-1}\left(\hat{F}_{ij}+\Lambda_{ij}\right)U,
\end{align}
where
\begin{align}
\Lambda_{ij}=\Omega_{[i|k|}x^{k}\left[\hat{A}_{j]},U\partial_{-}U^{-1}\right].
\end{align}
The anti-self-duality constraint is then preserved if
\begin{align}
\Lambda_{ij}=-\star\Lambda_{ij}.
\end{align}
It is unclear if this constraint has any nontrivial solutions. It would therefore be interesting to see if the gauge group embedding can be implemented more naturally by generalising the worldline representation in \eqref{eq: Phi general solution}.

Finally, as a simple example, let us consider the boosted version of the static
solution. In particular, take $\mu$ to be constant
\begin{align}
\mu(\tau)=\frac{\rho^{2}}{4\pi R} \ , 
\end{align}
but allow it to move in the $x_{4}$ direction with velocity $v$: 
\begin{align}
y^{-}(\tau)=\tau \ , \ y_{1}=y_{2}=y_{3}=0 \ , \ y_{4}=v\tau \ .
\end{align}
Further choosing $\Omega_{ij}=-R^{-1}\bar{\eta}^2_{ij}$ for concreteness, we find the covariant distance
\begin{align}
z(x,y)=x^- - \tau +  \frac{i}{4R}|\vec x |^2 + \frac{1}{2R}\left(x_{2}-ix_{4}\right)v\tau+\frac{i}{4R}v^{2}\tau^{2}\ ,
\end{align}
and the integral in \eqref{eq: Phi general single particle} gives 
\begin{align}
\Phi=1+\frac{\rho^{2}R\left(2\left(2R-vx_{2}\right)+2{\rm Re}\left(\frac{\left(2R-vx_{2}\right)\left(2R-vx_{2}+ivx_{4}\right)-2iRv^{2}x^{-}}{\sqrt{4R^{2}-4iRv\left(vx^{-}-x_{4}-ix_{2}\right)+v^{2}\left(\left(x_{2}-ix_{4}\right)^{2}+\left|\vec{x}\right|^{2}\right)}}\right)\right)}{4Rv\left(Rvx^{-}+vx_{2}x_{4}-2Rx_{4}\right)+\left(vx_{2}-2R\right)^{2}\left|\vec{x}\right|^{2}} \ .
\end{align}
This solution looks complicated, but in the limit where the velocity
goes to zero we recover the usual static solution:
\begin{align}
\lim_{v\rightarrow0}\Phi=1+\frac{\rho^{2}}{\left|\vec{x}\right|^{2}} \ .
\end{align}
Moreover, in the $R\to\infty$ ({\it i.e.}\ $\Omega_{ij}\to 0$) limit we recover a boosted version of the above
solution: 
\begin{align}
\lim_{R\rightarrow\infty}\Phi=1+\frac{\rho^{2}}{x_{1}^{2}+x_{2}^{2}+x_{3}^{2}+\left(x_{4}-vx^{-}\right)^{2}} \ .
\end{align}
Indeed, we later show in Section \ref{subsec: DLCQ} that we generally reproduce the usual 't Hooft form solutions in the $R\to\infty$ limit, with moduli that are allowed to vary arbitrarily with $x^-$.

\section{Gauge Topology and Instantons}\label{sec: gauge field topology}

We have already begun thinking of the solutions for $\hat{A}_i$ corresponding to $\Phi$ of the form (\ref{eq: Phi general solution}) as anti-instanton particles of the gauge field $A=(A_-, A_i)$. Let us now justify this.

We first return to the case of a single instanton centred at the origin, as given by $\Phi$ in (\ref{eq: Phi spherically symmetric solution}). Here, we will recover almost all of the important qualitative properties of the much more general solution (\ref{eq: Phi general solution}), while avoiding many of the more technical details.

We then generalise our analysis, and show that the solution (\ref{eq: Phi general solution}) describes an arbitrary number of anti-instanton particles, generically travelling between points at which they are created and annihilated.

\subsection{The Single, Spherically Symmetric Anti-Instanton}

Let us consider again the spherically-symmetric solution for $\hat{A}_i$ given by
\begin{align}
  \Phi[\mu] = 1+ \int_{-\infty}^\infty \frac{\mu(\tau)}{|\tau-z|^2} d\tau = 1+\frac{1}{z-\bar{z}} \int_{-\infty}^\infty d\tau\, \mu(\tau) \left(\frac{1}{\tau-z}-\frac{1}{\tau-\bar{z}}\right) \ ,
  \label{eq: Phi for single instanton at origin}
\end{align}
where recall the shorthand $z=z(x,0)=x^- + \frac{i}{4R}|\vec  x|^2$. This corresponds to a choice of $y^-(\tau)=\tau$ and $y^i(\tau)=0$ in the more general solution (\ref{eq: Phi general single particle}).

It is clear then that for suitable behaviour of $\mu(\tau)$ at large $\tau$ ({\it i.e.}\ that it is bounded), the integral converges when $|\vec  x|>0$, and so $\Phi$ is regular away from the origin. Indeed, the solution we will be most interested in are those for which $\mu(\tau)$ has compact support. Conversely, $\Phi$ and hence $\hat{A}_i$ is singular at all points such that $x^i=0$ and $\mu(x^-)>0$.

Now let $\mathbb{R}^4_*\cong \mathbb{R}^4$ denote the spatial slice defined by fixing some $x^- = x^-_\ast$. We can then consider the total instanton flux through this slice, defined by
\begin{align}
  Q_* = \frac{1}{8\pi^2}\int_{\mathbb{R}^4_*} \text{tr}\left(F\wedge F\right)\ .
\end{align}
In our conventions, $F=dA-i A\wedge A$, and so we have locally $\tr(F\wedge F)=d\nu_3(A)$, where $\nu_3(A)=\text{tr}\left(A\wedge dA -\frac{2i}{3}A\wedge A\wedge A\right)$ is the Chern-Simons 3-form.

It may be that the gauge field $A$ is regular throughout $\mathbb{R}^4_*$.
Then, $Q_*$ reduces to an integral of $\nu_3 (A)$ over the 3-sphere at spatial infinity. However more generically the 't Hooft ansatz produces  
solutions where $A$ is singular at $x^i=0$, if $\mu(x^-_*)>0$. Hence, we generally have that $Q_*$ reduces to
\begin{align}
  Q_*= \frac{1}{8\pi^2}\int_{S^3_\infty} \nu_3(A) - \frac{1}{8\pi^2}\int_{S^3_0} \nu_3(A)
\end{align}
where $S^3_0$ and $S^3_\infty$ denotes 3-spheres around the origin and at spatial infinity, respectively. 

Our key result is that, for suitable boundary conditions on $A_-$, we have $Q_*=0$ if $\mu(x^-_*)=\dot{\mu}(x^-_*)=0$, and $Q_*=-1$ if $\mu(x^-_*)>0$. In more detail, we find that when $\mu(x^-_*)>0$, the leading order behaviour of $A_i$ near the origin matches that of a single $SU(2)$ anti-instanton in singular gauge centred at the origin. As such, the integral of $(1/8\pi^2)\,\nu_3(A)$ over $S^3_0$ is quantised in the integers; it is indeed simply equal to $1$. If instead $\mu(x^-_*)=\dot{\mu}(x^-_*)=0$, then $A_i$ is regular at the origin and so the contribution to $Q_*$ from $S^3_0$ vanishes.

Conversely, we show that under reasonable assumptions on $\mu(\tau)$ as $\tau\to\pm\infty$, the contribution from the integral at $S^3_\infty$ vanishes, for any $\mu(x^-_*)\ge 0$. There are a number of steps required to arrive at this result. Firstly, we consider how the asymptotic behaviour of $\hat{A}_i$ both near the origin and at infinity is dictated by that of $\Phi$. Secondly, we must translate these asymptotics to those of the gauge field $A_i$ rather than $\hat A_i$, for which we must additionally consider the asymptotic behaviour of $A_-$.

We note that the form of $\hat{A}_i$ involves both $\Phi$ and $\partial_- \Phi$. This however does not pose much of a computational complication, since provided $\mu(\tau)$ is bounded as $\tau\to\pm\infty$, we have\footnote{Note, this relation holds only for solutions with $y^-=\tau$ and $y^i$ constant.} $\partial_-\Phi[\mu]=\Phi[ \dot \mu ]-1$, where $\dot \mu = d\mu/d\tau$. Thus, we have
\begin{align}
  \hat{A}_i (x) = -\eta^a_{ij} \sigma^a x^j \Phi[\mu]^{-1}  \frac{\partial\Phi[\mu]}{\partial|\vec x|^2} + \frac{1}{4} \eta^a_{ij} \sigma^a \Omega_{jk} x^k \Phi[\mu]^{-1} \left(\Phi[\dot{\mu}]-1\right)\ .
\end{align}
Let us now consider the behaviour of $\Phi[\mu]$, both as $|\vec x|\to\infty$ and $|\vec x|\to 0$. Firstly, it is immediate that as $|\vec x|\to \infty$, we have $\Phi[\mu]=1+\mathcal{O}(|\vec x|^{-2})$. However, the corresponding leading order behaviour of $\hat{A}_i$ requires that we know the next-to-leading-order behaviour of $\Phi$, which in turn depends subtly on the global properties of the function $\mu(\tau)$. 

First suppose that the integral of $\mu(\tau)$ over $\tau\in\mathbb{R}$ converges, so that in particular $\mu(\pm\infty)=0$.\footnote{Note that this is a sufficient but not necessary condition for $\Phi$ to be finite away from the worldline, which requires only that $\mu(\tau)$ is bounded as $\tau\to \pm\infty$.} Then, in the limit $|\vec x|^2\to \infty$, we have
\begin{align}\label{muint}
  \Phi[\mu] = 1 + \frac{16R^2}{|\vec x|^4} \left(\int_{-\infty}^\infty d\tau\, \mu(\tau)\right) + \mathcal{O}\left(|\vec x|^{-6}\right)\ .
\end{align}
More generally however we may consider profiles for $\mu(\tau)$ such that the limits $\mu(\pm\infty)=\lim_{\tau\to\pm\infty}\mu(\tau)$ exist but may be non-zero. Such choices will still give rise to finite $\Phi$ away from worldlines, but now the behaviour as $|\vec x|\to\infty$ is adjusted. We find\footnote{This can be seen by writing $\mu = \frac12(\mu(\infty)+\mu(-\infty))+\tfrac12(\mu(\infty)-\mu(-\infty))\tanh(\tau)+\mu_0(\tau)$  where $\mu_0$ has a finite integral, performing the integral of the first two terms explicitly, and then using (\ref{muint}) for the $\mu_0$ contribution.}
\begin{align}
  \Phi[\mu] = 1 + \frac{2\pi R}{|\vec x|^2}\left(\mu(\infty)+\mu(-\infty)\right) + \mathcal{O}\left(|\vec x|^{-4}\right)\ ,
\end{align}
provided that $\mu$ converges to $\mu(\pm \infty)$ at least as quickly as $\tau^{-2}$ as $\tau\to\pm \infty$.

Next, we can investigate the behaviour of $\Phi[\mu]$ as $|\vec x|\to 0$. This is most easily seen by first considering the Fourier transform of the function $\mu(\tau)$,
\begin{align}
  \mu(\tau) &= \int_{-\infty}^\infty d\omega\, e^{i\omega\tau} \tilde{\mu}(\omega)	\ , 	\nonumber\\
  \tilde{\mu}(\omega) &= \frac{1}{2\pi}\int_{-\infty}^\infty d\tau\, e^{-i\omega\tau} \mu(\tau)	\ .
\end{align}
with reality of $\mu(\tau)$ implying $\overline{\tilde{\mu}(\omega)}=\tilde{\mu}(-\omega)$. Note, non-zero $\mu(\pm\infty)$ corresponds to allowing for $\delta$-function profiles for $\tilde{\mu}(\omega)$.

We then have for all $|\vec x|\neq 0$,
\begin{align}
  \int_{-\infty}^\infty d\tau\, \frac{\mu(\tau)}{\tau-z} = \int_{-\infty}^0 d\omega \,\tilde{\mu}(\omega) \int_{-\infty}^{\infty} d\tau\, \frac{e^{i\omega \tau}}{\tau-z} + \int_0^{\infty} d\omega \,\tilde{\mu}(\omega) \int_{-\infty}^{\infty} d\tau\, \frac{e^{i\omega \tau}}{\tau-z}\ .
\end{align}
Both integrals over $\tau$ can then be computed by a corresponding contour integral, with the contour closed in the lower half-plane for the former, and upper half-plane for the latter. Since $z$ lies in the upper half-plane, only the latter integral survives, and we have
\begin{align}
  \int_{-\infty}^\infty d\tau\, \frac{\mu(\tau)}{\tau-z} = 2\pi i \int_0^\infty d\omega\,\tilde{\mu}(\omega)  e^{i\omega z}\ ,
\end{align}
and hence,
\begin{align}
  \Phi[\mu] 	&= 1+\frac{4\pi R}{|\vec x|^2} \int_0^\infty d\omega \left(\tilde{\mu}(\omega)e^{i\omega z}+\overline{\tilde{\mu}(\omega)}e^{-i\omega \bar{z}}\right)		\nn\\
  				&= 1+ \frac{4\pi R}{|\vec x|^2}\left[ \int_{-\infty}^0 d\omega\, \tilde{\mu}(\omega) e^{i\omega \bar{z}} + \int^{\infty}_0 d\omega\, \tilde{\mu}(\omega) e^{i\omega z} \right]\ .
\end{align}
Therefore, we find that as we approach $|\vec x|\to 0$,
\begin{align}\label{smallx}
  \Phi[\mu] = \frac{4\pi R}{|\vec x|^2} \mu(x^-) + \mathcal{O}(1)\ .
\end{align}
With these results in hand, we are ready to write down the asymptotic behaviour of $\hat{A}_i$. Firstly, as $|\vec x|\to\infty$ we have
\begin{align}
  \hat{A}_i = \mathcal{O}\left( |\vec x|^{-3} \right) \ , 
\end{align}
assuming that $\lim_{\tau\to\pm\infty}\dot{\mu}(\tau)=0$. Note that this holds even if we allow $\mu(\pm\infty)$ to be non-zero.

Next, consider the limit $|\vec x|\to 0$. Then, if $\mu(x^-_*)>0$, we find
\begin{align}
  \hat{A}_i = \frac{1}{|\vec x|^2}\eta^a_{ij} x^j\sigma^a +\mathcal{O}\left(|\vec x|\right)\ .
\end{align}
In contrast, if $\mu(x^-_*)=0$ and $\dot{\mu}(x^-_*)=0$, then we have
\begin{align}
  \hat{A}_i = \mathcal{O}\left(|\vec x|\right)\ .
\end{align}
Before we can say anything about $Q_*$, we must finally determine the corresponding asymptotic behaviour of $A_i$, determined in terms of $\hat{A}_i$ and $A_-$ by
\begin{align}
  A_i = \hat{A}_i +\frac{1}{2}\Omega_{ij} x^j A_-\ .
\end{align}
Then, if $A_-$ dies away at least as quickly as $|\vec x|^{-3}$ as $|\vec x|\to\infty$, and is no more singular than $|\vec x|^{-1}$ as $|\vec x|\to 0$, then the leading order behaviour of $A_i$ in these limits if $\mu(x^-_*)>0$ is given by
\begin{align}
  	A_i &= \mathcal{O}\left( |\vec x|^{-2} \right) \ , &&\hspace{-20mm} \text{as }|\vec x|\to \infty 		\nn\\
  	A_i &= \frac{1}{|\vec x|^2}\eta^a_{ij} x^j\sigma^a +\mathcal{O}\left(|\vec x|\right)\ , \qquad  &&\hspace{-20mm}\text{as }|\vec x|\to 0 \ .
\end{align}
Hence, the contribution to $Q_*$ from the integral over $S^3_\infty$ vanishes. Conversely, the behaviour of $A_i$ near the origin is precisely that of a single $SU(2)$ anti-instanton in singular gauge, centred at the origin, and thus the resulting contribution to $Q_*$ is quantised in the integers. Indeed, we have that the Chern-Simons 3-form $\nu_3$ pulled back to $S^3_0$ is given by $\nu_3|_{S^3_0}=\left(4+\mathcal{O}(|\vec x|^{-2})\right)d\Omega_3$, where $d\Omega_3$ is the standard volume form on $S^3$. Hence, the contribution to $Q_*$ from the integral over $S^3_0$ is precisely $-1$, and so we find $Q_*=-1$.

We are of course free to make such a choice of boundary condition for $A_-$, in effect defining some refined subspace of the total configuration space in which we require $A_-$ sits. However, it is a priori not clear that this subspace intersects with the subspace of solutions to the classical equations of motion, and thus such a boundary condition may violate any straightforward variational principle in the theory. However we will see below that there are solutions for $A_-$ which leave the instanton number  of $A_i$ intact and curiously that there are also solutions which precisely cancel the divergent behaviour of $\hat{A}_i$ near the worldline, and hence have $Q_*=0$.



\subsection{Creation and Annihilation}\label{subsubsec: creation and annihilation}

We have found that the instanton flux $Q_*$ on a slice $\mathbb{R}^4_*$ of constant $x^-=x^-_*$ depends in a crucial way on whether $\mu(x^-_*)>0$ or $\mu(x^-_*)=0$. This is indicative of a singularity not just in the gauge field but in the field strength $F$ itself, located at points at the spatial origin at which $\mu(x^-)$ transitions from a zero to non-zero value.

We can understand this as follows. Suppose $\mu(\tau)>0$ on $\tau\in(\tau_1, \tau_2)$, $\tau_1<\tau_2$, and identically zero otherwise. We have then that the total instanton flux over a constant $x^-$ slice is $-1$ if $x^-\in (\tau_1, \tau_2)$, while it is zero for $x^-\in (-\infty,\tau_1)\cup (\tau_2,\infty)$. We can thus interpret the point $x_1=(\tau_1,\vec 0)$ as the location at which an anti-instanton is created, and $x_2=(\tau_2,\vec 0)$ as the point at which it is annihilated.

We can gain further insight into the behaviour of the gauge field at the transition points $x_1, x_2$ by considering the instanton charge over more general four-dimensional submanifolds. Define $\Omega_4 = (1/8\pi^2)\,\text{tr}\left(F\wedge F\right)$, and write $Q(S)=\int_S \Omega_4$ for some submanifold $S$, so that for instance $Q_*=Q(\mathbb{R}^4_*)$. Note, it is clear that away from $\vec x=\vec 0$, we have $d\Omega_4=0$. We in fact have that $d\Omega_4=0$ everywhere except for at the transition points. This is seen by considering the integral of $\Omega_4$ over generic Gaussian pillboxes. Consider in particular $Q(P)$ where $P$ is a cylinder whose top and bottom lie transverse to the line $\vec x=\vec 0$. If such a cylinder does not intersect the line $\vec x=\vec 0$, then $A$ is defined globally over $P$ and hence $Q(P)=0$. Suppose instead that $P$ does intersect the spatial origin, but that it does not contain a transition point (see Figure \ref{fig: trivial pillbox}).

 Then, $Q(P)$ reduces to a pair of integrals of $\omega_3(A)$ on the small 3-spheres surrounding the two points at which the origin intersects $P$, with a relative minus sign due to orientation. But these two contributions are equal, and thus $Q(P)=0$. This is then sufficient to ensure that $d\Omega_4=0$ everywhere away from transition points.
 
\begin{center}
\begin{minipage}{\textwidth}
\centering
\captionof{figure}{Pillbox integral with vanishing $Q$.}\label{fig: trivial pillbox}
\hspace{-8mm}\includegraphics[width=70mm]{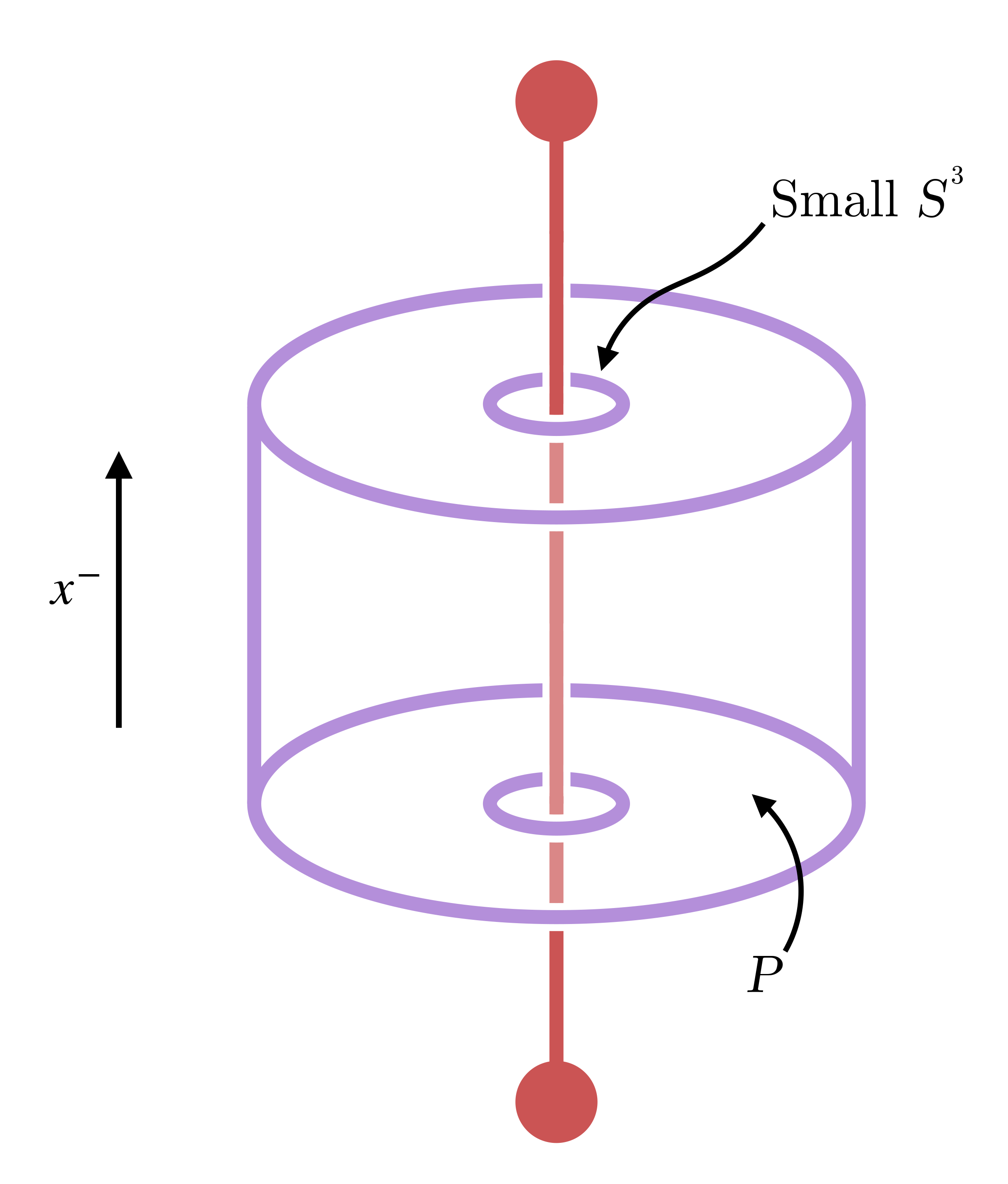}	
\end{minipage}
\end{center}
We can next consider $Q(S)$ for a generic submanifold $S$. The fact that $\Omega_4$ is closed away from the two transition points implies that $Q(S)$ is topological; we can smoothly deform $S$ without changing $Q(S)$, provided that such a deformation does not drag $S$ through a transition point. In particular, if $S$ doesn't contain either of $x_1, x_2$, then $S$ can be shrunk to a point and $Q(S)=0$. Suppose instead that $S$ contains $x_1$ (but not $x_2$). We can then smoothly deform $S$ to a cylinder of the type described previously (see Figure \ref{fig: creation pillbox}). It is clear then that $Q(S)$ receives a contribution of $-1$ from the top of the cylinder, but zero from the bottom, and hence $Q(S)=-1$. Similarly, for $S$ containing $x_2$, but not $x_1$, we have $Q(S)=+1$.
\begin{center}
\begin{minipage}{0.8\textwidth}
\centering
\vspace{1em}\captionof{figure}{Instanton charge around a creation point.}\label{fig: creation pillbox}\vspace{-1em}
\includegraphics[width=140mm]{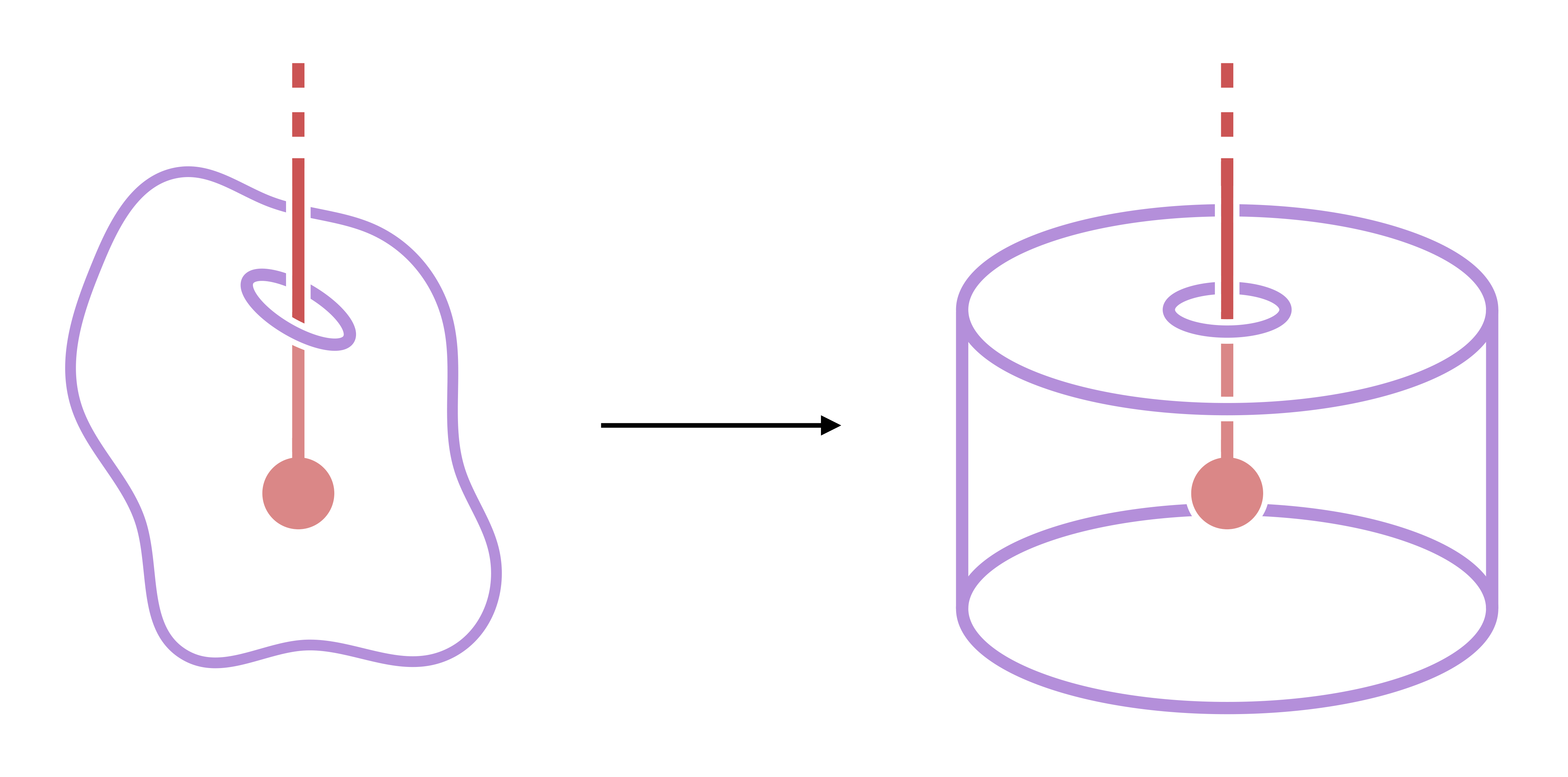}	
\end{minipage}
\end{center}
Indeed, we can consider $S$ to be some arbitrarily small 4-sphere about either a creation or annihilation point, for which we will still have $Q(S)=-1$ or $Q(S)=+1$, respectively. One can understand this configuration for $A$ more formally as the connection on a gauge bundle not over $\mathbb{R}^5$ but instead over $\mathbb{R}^5\setminus \{x_1, x_2\}$, whose non-trivialities are characterised by the integral of the second Chern class---{\it i.e.}\ $Q(S)$---over such small 4-spheres. Alternatively, we can extend such configurations to include the points $x_1,x_2$, provided we allow for the behaviour $d\Omega_4 = d^5x\left(\delta^{(5)}(x-x_2) - \delta^{(5)}(x-x_1)\right)$.

Note finally that this analysis generalises trivially to the case that $\mu$ varies from zero to non-zero and back not once but a number of times. Such a configuration describes an anti-instanton being created then annihilated, followed by another being created then annihilated, and so on as in example (\ref{wavey}).

\subsection{Moving Away from the Origin}

This analysis generalises easily to describe a single, static anti-instanton sitting not necessarily at $x^i=0$ but at a generic constant worldline $x^i=y^i$. This is not immediate, due to the unconventional translational symmetries present in the theories from which the constraint $\mathcal{F}_{ij}+\ast \mathcal{F}_{ij}=0$ arises, but is nonetheless not much more work to show. The $\Phi$ giving rise to such an anti-instanton can be written as
\begin{align}
  \Phi = 1+ \int d\tau \frac{\mu(\tau)}{z(x,y(\tau))\bar{z}(x,y(\tau))}\ ,
\end{align}
for $y(\tau)=(\tau, y^i)$ with constant $y^i\in\mathbb{R}$. More explicitly, we have
\begin{align}
  \Phi = 1+ \int_{-\infty}^\infty \frac{\mu(\tau)}{|\tau-z|^2} d\tau = 1+\frac{1}{z-\bar{z}} \int_{-\infty}^\infty d\tau\, \mu(\tau) \left(\frac{1}{\tau-z}-\frac{1}{\tau-\bar{z}}\right) \ ,
\end{align}
as in (\ref{eq: Phi for single instanton at origin}), except now $z=z(x,(0,y^i))=x^- +\frac{1}{2}\Omega_{ij} x^i y^j +  \frac{i}{4R}(x^i-y^i) (x^i-y^i)$. The important point, however, is that since the instanton is still static, $z$ is independent of $\tau$, and hence we can proceed identically as before.

We are again interested in $Q_*$, the instanton charge on a slice of constant $x^-_*$. In particular, the large $|\vec x - \vec y|\sim |\vec x|$ behaviour is such that $Q_*$ receives no contribution from spatial infinity, for suitable behaviour of $A_-$. Conversely, we can assess the behaviour near the wordline $x^i\to y^i$ by Fourier transform of $\mu(\tau)$, which gives us
\begin{align}
  \Phi	&= 1+ \frac{4\pi R}{|\vec x-\vec y|^2}\left[ \int_{-\infty}^0 d\omega\, \tilde{\mu}(\omega) e^{i\omega \bar{z}} + \int^{\infty}_0 d\omega\, \tilde{\mu}(\omega) e^{i\omega z} \right]\ ,
\end{align}
and hence as $|\vec x - \vec y|\to 0$,
\begin{align}
  \Phi &=\frac{4\pi R}{|\vec x - \vec y|^2}\mu\hspace{-0.5mm}\left(x^-+\tfrac{1}{2}\Omega_{ij} x^i y^j\right) + \mathcal{O}(1) 	\nn\\
  &=\frac{4\pi R}{|\vec x - \vec y|^2}\mu\hspace{-0.5mm}\left(x^-\right) + \mathcal{O}(|\vec x - \vec y|^{-1}) \ .
  \label{eq: single, static asymptotics away from origin}	
\end{align}
This is then enough to ensure that $\hat{A}_i$ and, for suitable boundary conditions for $A_-$, the gauge field $A_i$ behaves near the worldline precisely like a single anti-instanton in singular gauge, provided that $
\mu(x^-_*)>0$. If this is indeed the case, then $Q_*=-1$. If however $\mu(x^-_*)=0$ and $\dot{\mu}(x^-_*)=0$, we have $Q_*=0$. Indeed, the interpretation of transition points between these regions as creation and annihilation points, each carrying non-zero instanton charge on surrounding 4-spheres, generalises in the obvious way.

\subsection{General Worldlines}

We have seen that in the case that $y^i(\tau)$ is constant and $A_-$ regular, the resulting gauge field $A_i$ describes an anti-instanton sitting at $x^- = y^i$, that is created whenever $\mu$ transitions from a zero to non-zero value, and then annihilated when it returns to zero. These transition points are then special points in the spacetime, carrying non-zero instanton charge.

This interpretation extends in the natural way to the more general form of $\Phi$,
\begin{align}
  \Phi(x) = 1+ \sum_{A=1}^N\int d\tau_A \frac{\mu_A(\tau_A)}{z(x,y_A(\tau_A))\bar{z}(x,y_A(\tau_A))}\ ,
\end{align}
where we can assume without loss of generality that each of the $\mu_A(\tau_A)$ is strictly non-zero on some open interval $(\tau_{A,1}, \tau_{A,2})\subseteq \mathbb{R}$, and otherwise identically zero. Then, $\Phi(x)$ is regular throughout $\mathbb{R}^5$, except along curves defined by $y_A(\tau)=(y^-_A(\tau),y^i_A(\tau))$ for $\tau\in (\tau_{A,1}, \tau_{A,2})$, at which it is singular. If we further suppose that each of these curves extends in the $x^-$ direction without turning---more precisely that each of the $y^-_A(\tau)$ is a strictly monotonic function, which we are free to take as strictly increasing\footnote{This is because if we have a worldline with $y^-$ strictly decreasing, we can simply reparameterise $\tau\to-\tau$.}---then the resulting gauge field $A$ describes $N$ anti-instantons. Each is created at $y_A(\tau_{A,1})$, follows the worldline $y_A(\tau_A)$, and then is annihilated at $y_A(\tau_{A,2})$. 

To see this, let us first for simplicity of notation restrict our attention to the case of a single monotonic worldline and, as we did in the spherically symmetric case, consider the asymptotic behaviour of $\Phi$ at fixed $x^-$ as we approach the worldline. We are once again really interested in the resulting asymptotics of $\hat{A}_i$ which, for suitable boundary conditions on $A_-$, dictate the instanton charge $Q_*$ as measured over the slice $\mathbb{R}^4_*$ at constant $x^- = x^-_*$. We have
\begin{align}
  \Phi(x) = 1+ \int d\tau \frac{\mu(\tau)}{z(x,y(\tau))\bar{z}(x,y(\tau))}\ ,
\end{align}
where we have implicitly used the monotonicity of $y^-$ to reparameterise the worldline such that $y(\tau)=(\tau, y^i(\tau))$. The function $\mu(\tau)$ is strictly non-zero on $(\tau_1, \tau_2)\subseteq \mathbb{R}$, and identically zero otherwise.

Let us first fix $x^-=x^-_*$ such that $\mathbb{R}^4_*$ does not intersect the worldline, {\it i.e.}\ $x^-_*\notin [\tau_{1},\tau_{2}]$. Then, $\Phi$ is perfectly regular throughout $\mathbb{R}^4_*$, and dies away sufficiently fast as $|\vec x|\to \infty$ to ensure that for suitable behaviour of $A_-$, we have $Q_*=0$.

Suppose instead that $\mathbb{R}^4_*$ cuts through the interior of the worldline, that is $x^-_*\in (\tau^-,\tau^+)$. Then, as we approach $x^i \to y^i(x^-_*)$, the integral becomes increasingly divergent, with the dominant contribution from a neighbourhood of $\tau=x^-_*$. In this neighbourhood, we can write $y^i(\tau) = y^i(x^-_*)+\mathcal{O}\left(\tau- x^-_*\right)$, and thus as $|\vec x - \vec y(x^-_*)|\to 0$,
\begin{align}
  \Phi(x) &\sim 1+ \int d\tau \frac{\mu(\tau)}{z(x,(\tau,y^i(x^-_*)))\bar{z}(x,(\tau,y^i(x^-_*)))} \nn\\
  &\sim \frac{4\pi R}{|\vec x - \vec y (x^-_*) |} \mu(x^-_*)\ ,
\end{align}
which follows from (\ref{eq: single, static asymptotics away from origin}). Conversely, noting the sufficiently small behaviour as $|\vec x|\to\infty$, and for suitable behaviour of $A_-$, we find $Q_*=-1$.

These asymptotics then generalise to the case of $N$ monotonic worldlines, provided they do not intersect. Indeed, by generalising the arguments of Section \ref{subsubsec: creation and annihilation}, we can learn how to read off the value of $Q(S)$ for $S$ a 4-dimensional submanifold that does not pass through a transition point.

This can be summarised as follows. Suppose we have $N$ monotonic, disjoint worldlines, and let $S$ be some 4-dimensional submanifold that does not pass through any of the creation or annihilation points. Each anti-instanton is created at a point $x=y_A(\tau_{A,1})$ and annihilated at a point $x=y_A(\tau_{A,2})$, with $\tau_{A,2}> \tau_{A,1}$ and hence $y^-(\tau_{A,2})>y^-(\tau_{A,1})$. Thus, the $x^-$ direction defines an intrinsic direction of each worldline. Then, each time a worldline passes in this direction through\footnote{We assume that the intersection of $S$ and the set of all worldlines is a set of disjoint points in $\mathbb{R}^5$.} $S$ `upwards' in a right-handed sense, $Q(S)$ receives a contribution of $-1$, while each time it passes through `downwards', we pick up a $+1$. See Figure \ref{fig: multi disjoint} for an illustrative example. In particular, for $S$ a small 4-sphere surrounding a creation point, $Q(S)=-1$, while around an annihilation point, $Q(S)=+1$.

\begin{center}
\begin{minipage}{\textwidth}
\centering
\begin{minipage}{0.8\textwidth}
\captionof{figure}{The instanton charge $Q(S)=-1$ for a closed 4-dimensional surface $S$, which intersects some disjoint worldlines. Let $M$ be the region in $\mathbb{R}^5$ enclosed by $S$. Then, the two leftmost worldlines are those of anti-instantons created in $M$, and annihilated outside. The third describes an anti-instanton created outside $M$, and annihilated inside. The final, rightmost anti-instanton is both created \textit{and} annihilated outside $M$, but has worldline that nonetheless passes through $M$.}\label{fig: multi disjoint}
\end{minipage}
\includegraphics[width=140mm]{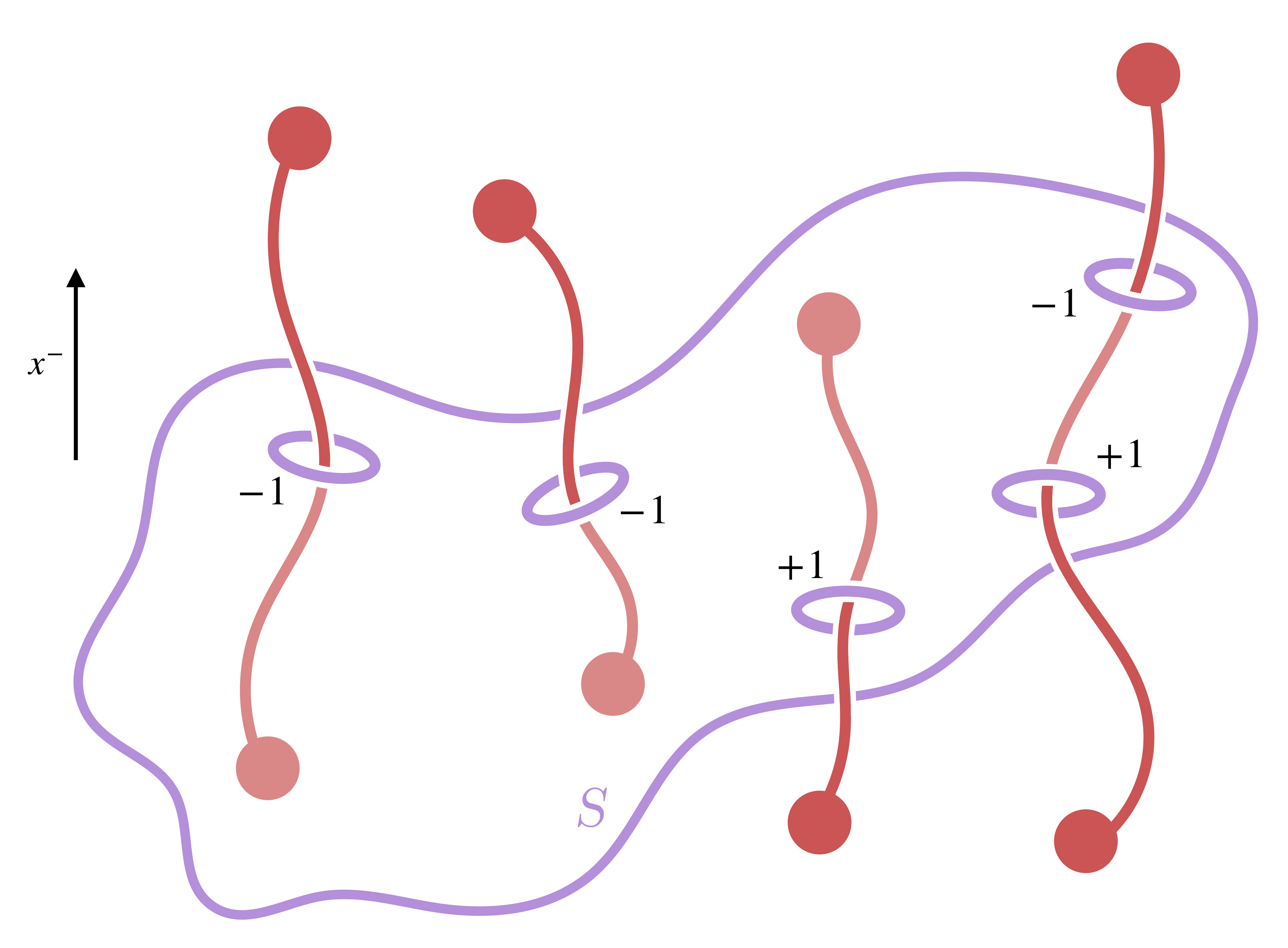}	
\end{minipage}
\end{center}

\subsection{Intersections, Turning Points, and Graphs}

Once we allow for general worldlines, there are a number of interesting additional features our worldlines may have that were not present in the static case. These give rise to an extended space of possible worldline configurations. Then, using the asymptotics we've already found, we can show that such configurations include transition points with not only $Q=\pm 1$, but generic $Q\in\mathbb{Z}$. 

So suppose we once again start with the $N$-instanton solution,
\begin{align}
  \Phi(x) = 1+ \sum_{A=1}^N\int d\tau_A \frac{\mu_A(\tau_A)}{z(x,y_A(\tau_A))\bar{z}(x,y_A(\tau_A))}\ ,
\end{align}
but let us go beyond the choice of monotonic, disjoint worldlines. First, we can consider what happens when a pair of worldlines intersects at one or more isolated points. By suitably splitting up any worldlines that intersect in their interiors into smaller worldlines joined end-to-end, we can reformulate this configuration as a set of worldlines that are disjoint in their interiors, but may share creation and annihilation points.

We could also suppose that one or more worldline has a turning point: a point at which the corresponding $\dot{y}^-(\tau)$ flips sign, and the worldline turns around. However, so long as we restrict our focus to worldlines for which $\dot{y}^-=0$ only at isolated points, we can once again split up such worldlines into into smaller sections, on each of which $y^-$ is monotonic. For example, a worldline with $y^-(\tau)=\tau(\tau-1)(\tau+1)$ and $\mu(\tau)$ non-zero for $\tau\in(-2,2)$ is split into three monotonic worldlines, forming a graph between four transition points (see Figure \ref{fig: turning points}). It is important here to remember that, due to our freedom to reparameterise, the only sense of `direction' for a worldline is that which corresponds to increasing $x^-$. Thus, there is no sense in which such a split-up worldline `remembers' it was once a single worldline with turning points.
\begin{center}
\begin{minipage}{0.8\textwidth}
\centering
\captionof{figure}{Splitting of a turning worldline into several monotonic pieces}\label{fig: turning points}
\includegraphics[width=80mm]{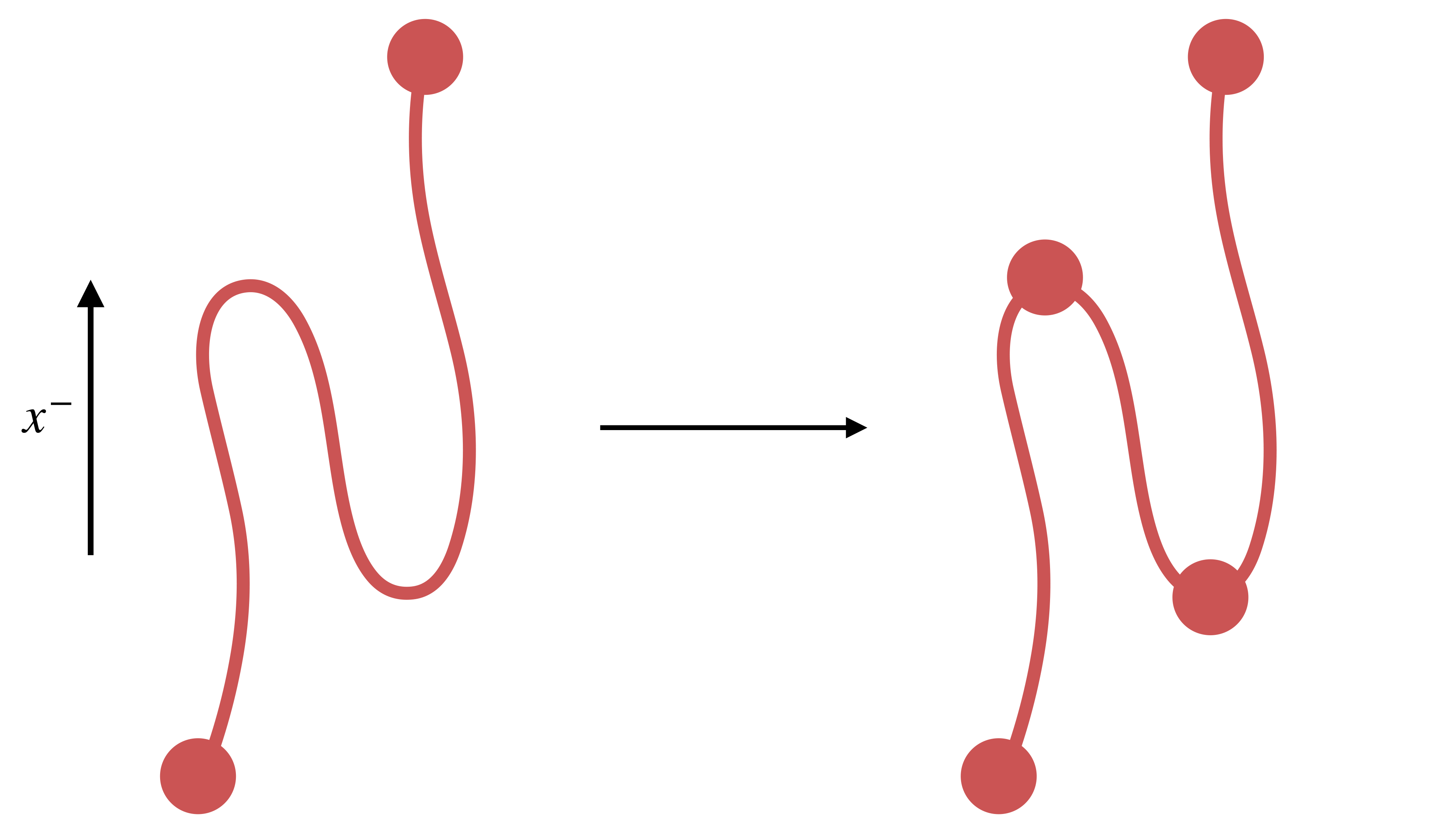}	
\end{minipage}
\end{center}
So, we are lead to a more general set-up:\ we still have $N$ monotonic worldlines, but now they are allowed to share beginnings and ends. A general worldline configuration is then a graph, whose nodes are a set of transition points, and whose edges are a set of monotonic worldlines. However, despite this generalisation, our rules for computing $Q(S)$ for some generic 4-dimensional surface $S$ carry over straightforwardly, as they care only about the asymptotic behaviour of the gauge field in a neighbourhood of the point at which a worldline intersects $S$. In particular, each time a worldline passes through $S$ `upwards' in a right-handed sense, $Q(S)$ receives a contribution of $-1$, while each time it passes through `downwards', we pick up a $+1$. See Figure \ref{fig: general graph} for an illustrative example.
 
\begin{center}
\begin{minipage}{0.8\textwidth}
\centering
\captionof{figure}{The instanton charge $Q(S)=+1$ on a closed surface $S$, amongst transition points joined with monotonic worldlines. The middle two transition points lie in the region enclosed by $S$, while the other two are outside.}\label{fig: general graph}\vspace{-1em}
\includegraphics[width=110mm]{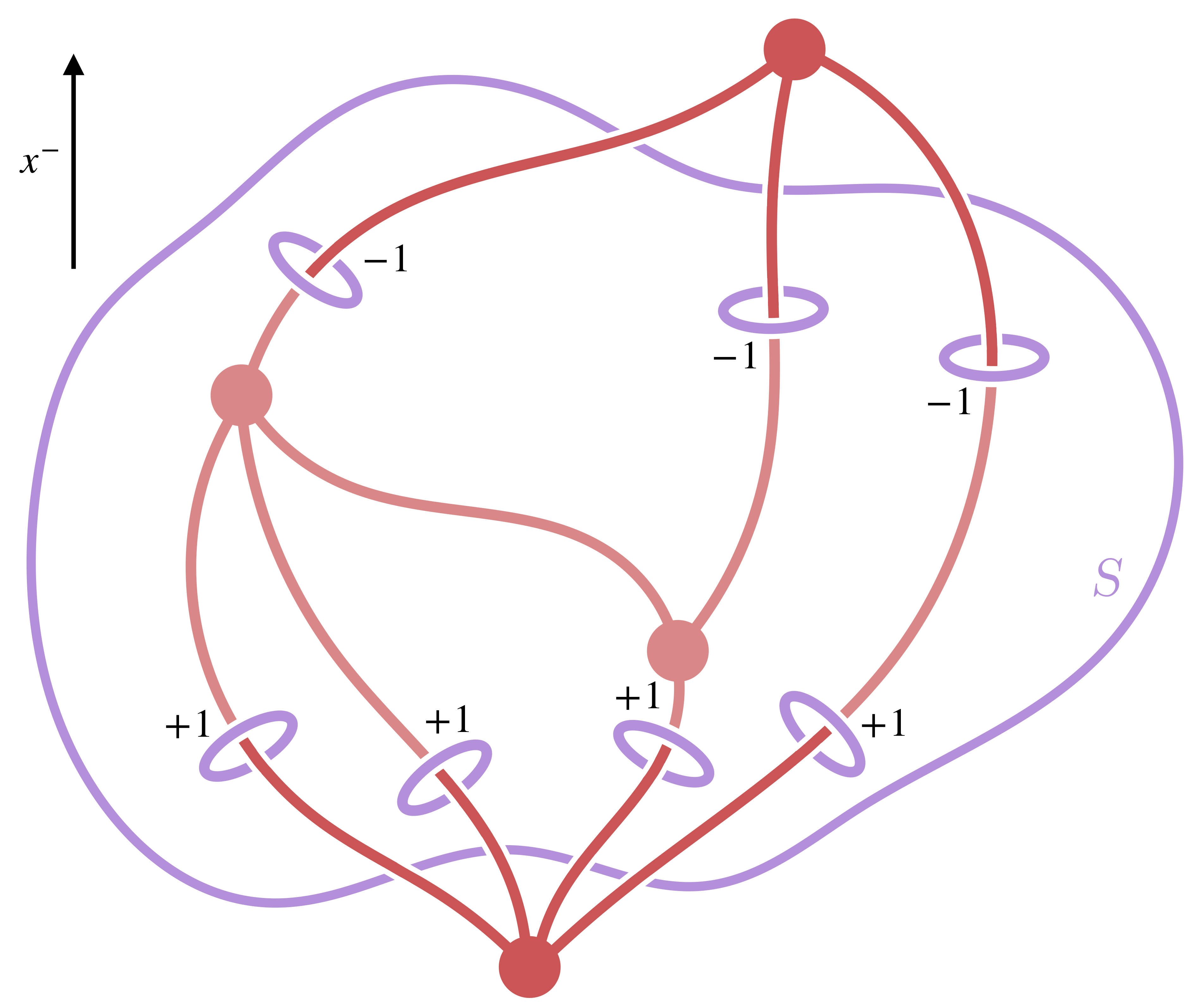}	
\end{minipage}
\end{center}
It is also interesting to ask what $Q(S)$ is when $S$ is a small 4-sphere surrounding some transition point $y\in\mathbb{R}^5$. It is given simply by the number of worldlines annihilated at $y$, minus the number created at $y$ (see Figure \ref{fig: general transition point}). In this way, we can construct configurations with arbitrary $Q\in\mathbb{Z}$.\\

Lastly let us make some comments. Firstly it might appear as though there is a little ambiguity in our analysis: given any graph of transition points and monotonic worldlines, we can always split any worldline into a pair of monotonic worldlines, in effect introducing a new transition point. However, such a point has $Q=0$ on a small 4-sphere surrounding it, and so therefore will not give rise to a singularity in $F$.

Secondly,
and we will discuss this in greater detail below, it is not hard to see that finiteness of the action (at least for $A_-=0$) requires that the positions $y^-$, $\vec y_A$ as well as $\mu_A$ are suitably well-behaved functions of $\tau_A$. In particular we find finite actions so long as their derivatives with respect to $\tau_A$ are bounded and vanish as $\tau_A\to\infty$.

\begin{center}
\begin{minipage}{0.8\textwidth}
\centering
\captionof{figure}{A general transition point, with $Q(S)=m-n$ on a small 4-sphere $S$ surrounding it.}\label{fig: general transition point}
\includegraphics[width=90mm]{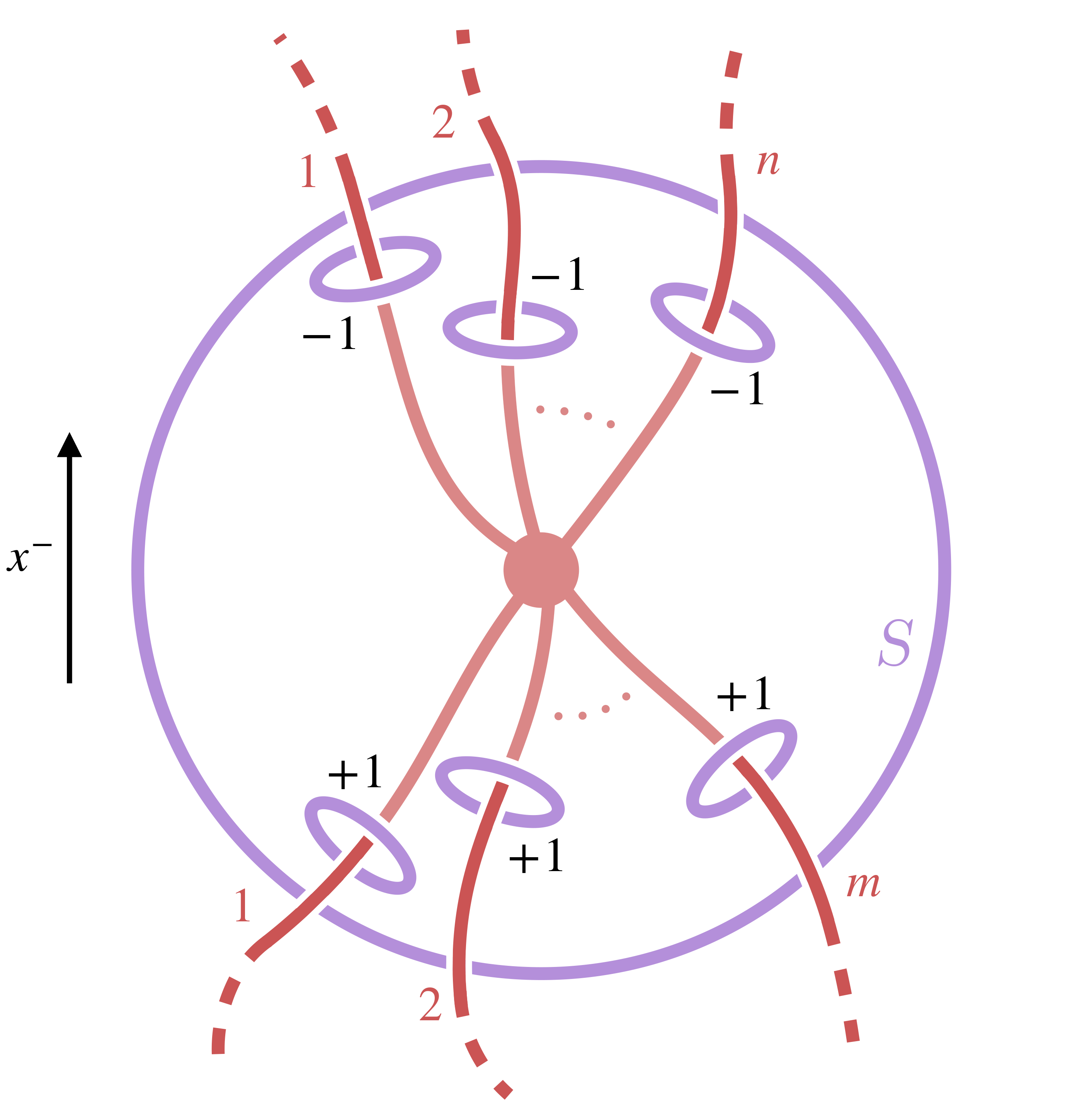}	
\end{minipage}
\end{center}

Finally, note that the interpretation of turning points as transition points, {\it i.e.}\ points at which $\tr(F\wedge F)$ is singular, is forced upon us. This is seen most simply by considering $Q(S)$ for $S$ a small 4-sphere surrounding such a point, which following our discussion gives $Q(S)=-2$ at a local minimum (the creation of two anti-instantons), or $Q(S)=+2$ at a local maximum (the annihilation of two anti-instantons). 

It is instructive to look at a simple example. We can consider a worldline with $y^-(\tau)=\tau^2$ and $y^i(0)=0$, which has a local minimum when $\tau=0$ at the spacetime origin $x=(0,\vec{0})$. Suppose we attempt to calculate $Q(S)$ for $S\cong \mathbb{R}^4$ the spatial slice at constant $x^-=0$, which includes the turning point. As we approach $|\vec x|\to 0$, we have
  \begin{align}
	\Phi &\sim 1 +\int_{-\infty}^\infty d\tau  \frac{\mu(\tau) }{|x^--\tau^2+\tfrac{i}{4R}|\vec x|^2|^2} \nonumber\\
	& = 1 +\int_0^\infty d\tau'\frac{1}{2\sqrt{\tau'}}  \frac{\mu(\sqrt{\tau'})+\mu(-\sqrt{\tau'}) }{|x^--\tau'+\tfrac{i}{4R}|\vec x|^2|^2}		\nn\\
	&\sim \frac{2\pi R}{|\vec x|^2} \lim_{x^- \to 0^+}\left(\frac{\mu(\sqrt{x^-})+\mu(-\sqrt{x^-})}{\sqrt{x^-}}\right) \ .
 \end{align}
If $\mu(0)\neq 0$, this limit does not exist and $\Phi$ is too divergent at the origin, leading to an infinite action.  Thus for $\mu(0)\neq 0$, $Q(S)$ is ill-defined and $x=(0,\vec{0})$ is a transition point. Conversely, if $\mu(0)=0$ (but is still non-zero elsewhere), then $x=(0,\vec{0})$ is naturally thought of as the start of two distinct worldlines. In either case, we find that the turning point is indeed a transition point. Note further from this form of $\Phi$ that boundedness of $\partial_-\Phi$, needed for a finite action, also requires that $\dot\mu(0)=0$.

\subsection{Constraints on $Q$}

Despite our freedom in defining worldline configurations with arbitrary $Q\in\mathbb{Z}$, the global structure of such configurations nonetheless give rise to interesting constraints on the instanton charges of each transition point.

Suppose we have a worldline configuration with transition points $x_a$, $a=1,\dots, m$, with $x_a^-<x_{a+1}^-$ for each $a=1,\dots, m-1$, and suppose further that all worldlines are created or annihilated at one of these transition points, as opposed to any escaping to or from infinity. For each $a$, define $n_a=Q(S^4_a)\in\mathbb{Z}$, where $S^4_a$ is a small 4-sphere surrounding $x_a$.

By considering $Q(S)$ for some $S$ surrounding every $x_a$, we have $\sum_a n_a=0$. Indeed, we can also consider $\sum_{a=1}^r n_a$ for some $r=1,\dots, m-1$. Then, we have,
\begin{align}
  \sum_{a=1}^r n_a = \sum_{a=1}^r Q(S^4_a) = Q\left(\cup_{a=1}^r S^4_a\right) = Q\left(M_r\right) \ .  
  \label{eq: multi point deformation}
\end{align}
Here, we smoothly deformed the disjoint union of small 4-spheres into some closed 4-manifold $M_r$  that encloses the $x_1,\dots, x_r$, but not the $x_{r+1},\dots, x_m$. Crucially, this deformation can always be done without the 4-manifold passing through any transition points, and as such, $Q$ is invariant under the deformation.

Then, we have $Q(S_r)\le 0$. To see this, note that we can further deform $M_r$ to a cylinder $P_r$, with top and bottom at $x^-=y^-_1, y^-_2$ respectively, again without passing through any transition points. We have in particular that the top of $P_r$ lies somewhere between $x_r$ and $r_{r+1}$ ({\it i.e.}\ $x^-_r<y^-_2<x^-_{r+1}$), while its bottom lies below all transition points ({\it i.e.}\ $y^-_1<x^-_1$). We can further take the radius of the cylinder to be sufficiently large that the only points at which a worldline passes through $P_r$ is on its top. Hence, following the rules of the previous section, and taking note of the orientation of $P_r$ as inherited from that of $M_r$, we have $Q(M_r)=Q(P_r)\le 0$.

Figure \ref{fig: multi point deformation} provides a schematic of this calculation for $m=4$ and $r=2$, in particular demonstrating the continuous deformation of $S^4_1\cup S^4_2$ into $M_2$ and then into $P_2$.

Therefore, using $Q(S_r)\le 0$, we find
\begin{align}
  \sum_{a=1}^r n_a\le 0 \quad \text{for all }r=1,\dots m \ , 
  \label{eq: graph rule}
\end{align}
or equivalently, $ \sum_{a=r}^m n_a\ge 0$ for all $r=1,\dots, m$. 

Let us finally suppose further that the graph of transition points and worldlines is \textit{connected}. In the above discussion, this then implies that $Q(S_r)=Q(P)<0$ for all $r=1,\dots, m-1$, {\it i.e.}\ that the bound (\ref{eq: graph rule}) is saturated only for $r=m$. Therefore, we have the strengthened statement,
\begin{align}
	\sum_{a=1}^r n_a &< 0\qquad \text{for all }r=1,\dots, m-1	
  \label{eq: connected graph rule}
\end{align}
or equivalently, $\sum_{a=r}^m n_a > 0$ for all $r=2,\dots, m$.

\begin{center}
\begin{minipage}{0.8\textwidth}
\centering
\captionof{figure}{A schematic showing the continuous deformation of a pair of small 4-spheres $S^4_1\cup S^4_2$, to a single surface $M_2$ that encloses $x_1$ and $x_2$, and finally to a cylinder $P_2$ that is only pierced by worldlines on its top. The arrows represent orientation. Indeed, we can read off $Q(S^4_1)+Q(S^4_2)=-3+1=-2=Q(M_2)=Q(P_2)$.}\label{fig: multi point deformation}
\includegraphics[width=100mm]{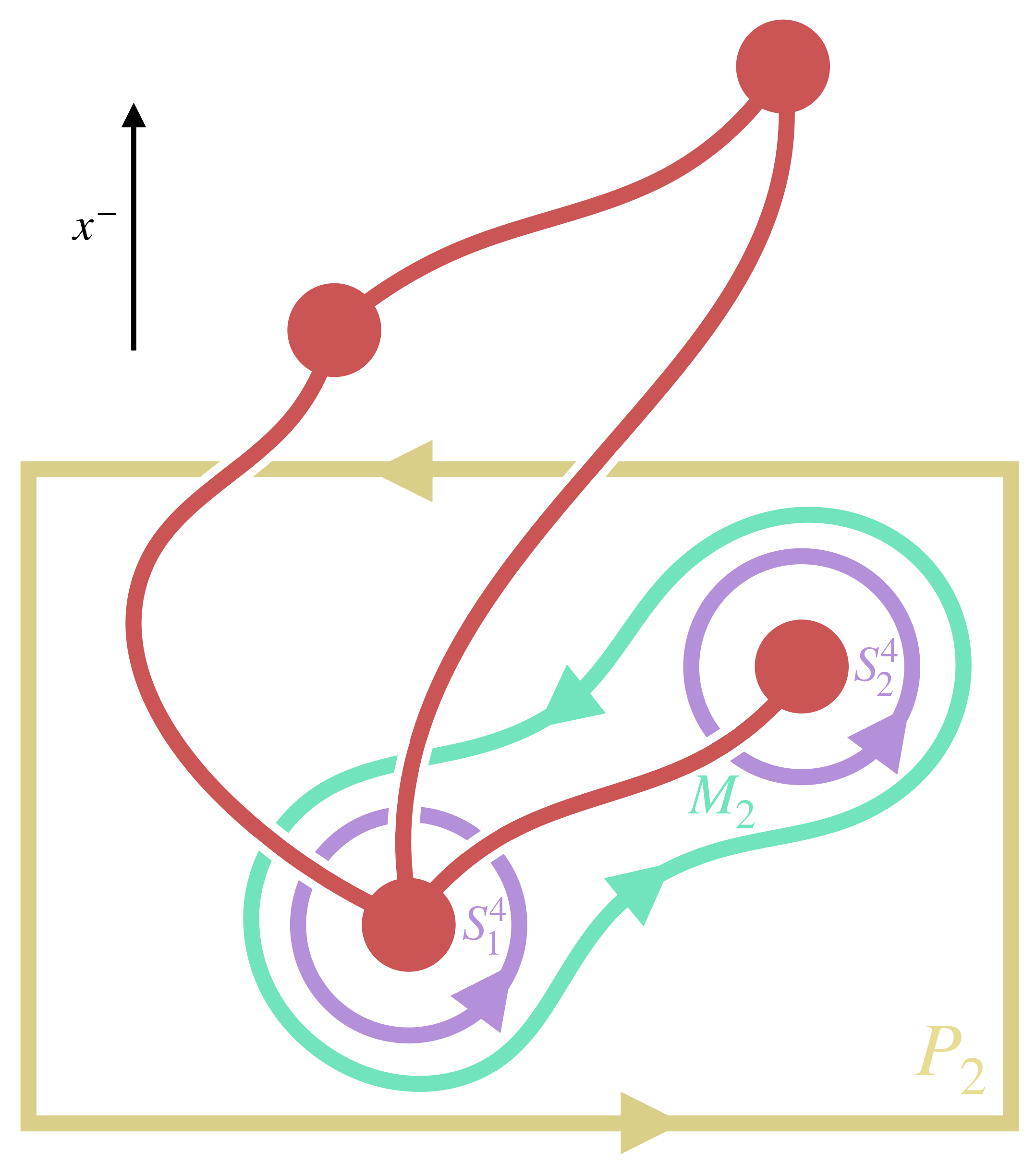}	
\end{minipage}
\end{center}

Each of these results (\ref{eq: graph rule}), (\ref{eq: connected graph rule}) is a straightforward application of the fact that on any spatial surface we will always have $Q({\mathbb R}^4_{x^-})\le 0$ for any $x^-$, provided there are no transition points lying precisely on ${\mathbb R}^4_{x^-}$. This is strengthened to $Q({\mathbb R}^4_{x^-})< 0$ in the case that the worldline graph is connected, and $x_1^-<x^-<x_m^-$.

Let us briefly compare these results to results on the correlation functions of $SU(1,3)$ theories \cite{Lambert:2020zdc}. One finds that the modes of a particular conformal compactification of a CFT on six-dimensional Minkowski space are precisely described by theories in five-dimensions with an $SU(1,3)$ spacetime symmetry. There is then hope that such a theory in fact captures the full spectrum of six-dimensional operators, through the inclusion of isolated points carrying non-trivial instanton charge on small spheres surrounding them \cite{US}. These are precisely the `transition points' of this paper. In a quantum treatment, such points are correspond to the insertion   of instanton operators as in \cite{Lambert:2014jna,Tachikawa:2015mha,Bergman:2016avc}.

One can then find further constraints on such a six-dimensional interpretation to work by dimensionally reducing correlation functions from six dimensions. In more detail, let $\mathcal{O}_n$ denote the $n\text{th}$ Kaluza-Klein mode of some six-dimensional operator $\mathcal{O}$, corresponding in the five-dimensional theory to a local operator dressed with an instanton operator of charge $n$. One can then find the  explicit form of the 2- and 3-point functions of the $\mathcal{O}_n$, as well as broader results on generic higher point functions. Strikingly, one finds that, at least at 2- and 3-points, the correlator\footnote{Here, the ordering of operators is fixed by a particular $i\epsilon$ prescription descended from six-dimensions.} $\left\langle \mathcal{O}_{n_m}(x_m) \dots \mathcal{O}_{n_2}(x_2) \mathcal{O}_{n_1}(x_1) \right\rangle$ is non-zero only if $\sum_{a=1}^r n_a < 0$ for each $r=1,\dots, m-1$, and for $r=m$ the sum vanishes.

Remarkably, this rule is precisely that which is satisfied for all connected worldline graphs (\ref{eq: connected graph rule}), which suggests that it should apply to higher-point correlators. It would be interesting to investigate if this holds for the dimensional reduction of 4-point correlators of protected operators in the 6d $(2,0)$ theory, which can be computed in the large-$N$ expansion \cite{Arutyunov:2002ff,Heslop:2004du,Rastelli:2017ymc,Heslop:2017sco,Chester:2018dga,Abl:2019jhh,Alday:2020lbp,Alday:2020tgi}.

 \section{Scalar Field Solutions}

Let us now examine the behaviour of the scalar fields on the constraint surface. In general there is always one scalar $\phi$ in the adjoint representation that comes from the tensor multiplet. However there can also be additional scalars $X^\alpha{}_m$ in an arbitrary representation coming from hyper-multiplets. Any such a scalar, which we denote by  $X$, appears in the action  through
\begin{align}
S_{scalar} =-\frac{1}{g^2_{YM}}\int dx^-d^4x \ \hat D_i X^\dag \hat D_i X\ .
\end{align}
Here $X$ is taken to be an any unitary representation $R$ of the gauge group (which we take to be $SU(2)$):
\begin{align}
\hat D_i X = \hat\partial_i X - i\hat A^a_iT_a(X)\ .	
\end{align}

First look at the classical equation of motion
\begin{align}
	\hat D_i\hat D _i X =0\ .
\end{align}
Smooth solutions to this equation are unique up to their behaviour at the boundary by a variation of the usual argument (note we    need the full five-dimensional integral here):
\begin{align}
 \int dx^-d^4x \hat D_i X^\dag \hat D _i X& = \int\hat \partial_i( X^\dag\hat D_i  X)-\int dx^-d^4x X^\dag \hat D_i\hat D _i X    \nonumber\\
 & = \oint X^\dag \hat D X\ .
\end{align}
Thus if $X$ vanishes on the boundary then \begin{align}
\hat D_iX	=0\ ,
\end{align}
everywhere and so $X=0$. In addition if $X$ is the difference   between two solutions which agree on the boundary then $X$ is also a solution but since it vanishes on the boundary the  two solutions must be equal everywhere.

Let us look more carefully at the boundary term. Since $\hat \partial_i$ contains derivatives in $x^-$ and $x^i$ we find
\begin{align}
 \oint X^\dag \hat D X & =    \int_{|\vec x|\to\infty} dx^-d\Omega_3 \, 	|\vec x|^2 x^i X^\dag \hat D_i X \nonumber\\
 &\qquad -\frac12\ \Omega_{ij} \int_{x^-\to\infty} d^4x \, x^jX^\dag \hat D_i X +\frac12\ \Omega_{ij} \int_{x^-\to-\infty} d^4x \, x^jX^\dag \hat D_i X \ ,\end{align}
where  $d\Omega_3$ is the volume element on a unit 3-sphere. Thus specifying the behaviour on the boundary means that we must specify the spatial behaviour in the form $X = X_{0} + X_1/|\vec x|^2+\dots $ but also the early and late values of $  X$ over all of ${\mathbb R^4}$. We will examine these terms in greater detail below.

In the case of the 't Hooft ansatz we can be quite explicit and 
compute
\begin{align}\label{DDX}
\hat D_i\hat D _i X =	\hat\partial_i\hat\partial_i X -\frac{C_R}{\Phi^2}\hat\partial_i\Phi \hat\partial_i\Phi X  + \frac{2i}{\Phi}\eta^a_{ij} \hat\partial_i\Phi T_a(\hat\partial_j X)\ ,
\end{align}
where we introduced $C_R$ as the quadratic Casimir of the representation:
\begin{align}
\sum_a T_aT_a = C_R{\mathbb I}	\ .
\end{align}
For example  in the adjoint representation $(T_a)_{bc}=-i\varepsilon_{abc}$ and hence $C_{adj}=2$, whereas for the fundamental representation $T_a=\tfrac12 \sigma_a$ and hence $C_{fund}=3/4$. More generally $C_R=s(s+1)$ with $s=0,\tfrac12,1,...$.
To solve this equation we impose the ansatz $X=X(\Phi(x^-,\vec x))$   so that the last term in (\ref{DDX}) vanishes. In this case we find
\begin{align}
\hat D_i\hat D _i X = \hat\partial_i\Phi \hat\partial_i\Phi \left(X'' - \frac{C_R}{\Phi^2} X\right) =0 	\ ,
\end{align}
where a prime denotes a derivative with respect to $\Phi$. Thus the  solutions to this take the form
\begin{align}
X = X_{0}\Phi^{\frac{1-\sqrt{1+4C_R}}{2}} + X'_{0}\Phi^{\frac{1+\sqrt{1+4C_R}}{2}}\ ,
\end{align}
for constant vectors $X_{0},X'_{0}$. 
However we want well behaved solutions at the poles of $\Phi$ and hence we find
\begin{align}
X &= X_{0}{\Phi^{-{\frac{\sqrt{1+4C_R}-1}{2}}}}\nonumber\\
&= X_{0}{\Phi^{-s}} \ .\end{align}
Note that there may exist other non-singular solutions since the boundary also contains pieces from $x^-\to\pm\infty$.

\section{Dynamics}\label{sec: dynamics}

In this section we will describe how to solve the equations of motion on the constraint surface and evaluate the action. To this end write the Bosonic part of the action as 
\begin{align}
S = \frac{1}{g^2_{YM}} \int dx^- d^4x\, \Big\{ \frac12 \tr \big(F_{-i}F_{-i})  + \frac{1}{2} \tr\big(\hat {F}_{ij}G^+_{ij} \big ) -\frac{1}{2} \tr \big(\hat D_{i} \phi\hat D_{i} \phi \big) -   \hat D_i {X}_{\alpha}{}^{m}\hat D_i X^{\alpha}{}_{m} \Big\}\ .
\end{align}
Varying with respect to $G^+_{ij},  A_i,  A_-, \phi$ and $X^\alpha{}_m$ respectively we find the equations of motion
\begin{align}
\hat{ F}_{ij} &= -\star \hat{ F}_{ij} \ , \nonumber\\
\hat D_j G^+_{ij}&= D_-F_{-i}  - i[\phi, \hat D_i\phi] - i[[{X}_{\alpha}{}^{m},\hat D_i X^{\alpha}{}_{m}]] \ , \nonumber\\
\frac12 \Omega_{ik}x^k\hat D_j G^+_{ij}&=D_iF_{-i}- \frac{i}{2} \Omega_{ik}x^k[\phi, \hat D_i\phi] - \frac{i}{2} \Omega_{ik}x^k[[{X}_{\alpha}{}^{m},\hat D_i X^{\alpha}{}_{m}]] \ , \nonumber\\
0&= \hat D_i \hat D_i \phi \ , \nonumber\\
0&= \hat D_i \hat D_i X^\alpha{}_m\ .
\end{align}
Here \begin{align}
 [[X_\alpha{}^m,\hat D_i X^\alpha{}_m]]	=\sum_a X_{\alpha }{}^mT_a(\hat D_i X^{\alpha }{}_m) T_a^{adj}\ ,
 \end{align}
  where $T_a$ are the $SU(2)$ generators for  the representation that $X^\alpha{}_m$ belongs to and $T_a^{adj}$ are the adjoint generators. 

We view the first equation as restricting the dynamics to the constraint surface defined by $\hat{ F}_{ij}=-\star \hat{ F}_{ij}$.
We can view the second equation as determining $G^+_{ij}$. However there is no need to explicitly solve for $G^+_{ij}$ as its contribution  to the action   will vanish on the constraint surface. 
Combining the second and third equations we simply find
\begin{align}\label{A-eq}
\hat D_i F_{-i} =\hat D_i \partial_-\hat A_i	- \hat D_i \hat D_i A_-	=0\ .
\end{align}
Here we find a scalar Laplace equation for $  A_-$ but now with a source. We can therefore find a unique solution for $  A_-$ for a given choice of boundary condition. Let us decompose 
\begin{align}
A_- = a_- + A'_-	 \ ,
\end{align}
where 
\begin{align}\label{A'-}
	\hat D_i\hat D_i a_-=\hat D_i\partial_- \hat A_i\ ,\qquad \hat D_i\hat D_i  A'_-=0\ ,
\end{align}
In particular we choose $a_-$ such that $a_-=0$ when $\partial_-\hat A_i=0$. Furthermore since $\partial_-\hat A_i\to 0$ on the boundaries we expect $a_-\to 0$ there  whereas  $\hat A'_-$ can be non-vanishing. Note that under a gauge transformation we require
 \begin{align}
 a_- &\to ig\partial_-g^{-1} + g a_- g^{-1}\ , \nonumber\\
 A'_- &\to g A'_- g^{-1}	\ ,
 \end{align}
 so we can think of $A_-'$ as an adjoint valued scalar which satisfies the same equation of motion as the scalar $\phi$, although it will have a different interpretation. Clearly if we start from a static ansatz with $a_-=A'_-=0$ can make $a_-$ non-zero by considering $x^-$-dependent gauge transformation while maintaining $A'_-=0$. In this sense we can think of $\hat D_i\hat D_i a_-=\hat D_i\partial_- \hat A_i$ as a gauge fixing condition.
   Thus we expect to find unique solutions for $a_-$ as well as $A'_-$, $\phi$ and $X^\alpha{}_m$ given their boundary values. 
 
We can now evaluate the action on the constraint surface to be
 %
\begin{align}\label{S}
S  
&= \frac{1}{g^2_{YM}} \int dx^- d^4x  \ \frac12 {\rm tr}(  \partial_- \hat A_i -  \hat D_i a_-\ -\hat D_i \hat A'_-)^2     -\frac{1}{2} \tr (\hat D_{i} \phi\hat D_{i} \phi \big ) -   \hat D_i {X}_{\alpha}{}^{m}\hat D_i X^{\alpha}{}_{m}\nonumber\\
 &= \frac{1}{g^2_{YM}} \int dx^-d^4x   \frac12 {\rm tr}(  \partial_- \hat A_i -  \hat D_i a_-)^2 - \frac{1}{g^2_{YM}}\oint  	\Big[  \tr \big( (\partial_- \hat A -  \hat D a_-)   A'_-  \big) -   \frac12 \tr ( A'_- \hat D A'_- )
	\Big]
\nonumber\\ &\qquad - \frac{1}{g^2_{YM}}\oint  	\Big[  \frac12\tr ( \phi  \hat D \phi \big ) +   {X}_{\alpha}{}^{m}\hat D   X^{\alpha}{}_{m} \Big] \ .
\end{align}
The first term gives an action for the gauge field $\hat A_i$. All the remaining terms are boundary contributions and as such depend on the choice of the asymptotic values of $A_-',\phi$ and $X^\alpha{}_m$ which are not fixed.

 Let us now discuss what this action looks like using the   't Hooft ansatz. 
 As seen above, for the scalars,  we take
\begin{align}\label{scalars}
 A'_- = A_{0-}\Phi^{-1}\ ,\qquad \phi = \phi_0\Phi^{-1}\ ,\qquad X^\alpha{}_m = X^\alpha_{0m}\Phi^{-s_m} \ ,
 \end{align}
from some constants $A_{0-},\phi_0\in su(2)$ and $X^\alpha_{0m}$ in the $SU(2)$ representation space of $X^{\alpha}{}_m$ with spin $s_m$.  These choices correspond to a specific set of boundary conditions where the fields approach constant values as $|\vec x|\to\infty$ whereas the  $x^-=\pm \infty$ behaviour is determined by $\Phi$.

However we need to determine $a_-$. This was required to solve $\hat D_i\hat D_i a_- = \hat D_i\partial_-\hat A_i$ such that it vanishes when $\partial_-\hat A_i=0$. This seems too complicated to do in general. However it is important to look at the solution near the instanton worldlines to check that they do not affect the original $A_i$ gauge field topology, as discussed in Section \ref{sec: gauge field topology}. For simplicity we can consider a static worldline  at $ x^i=0$ and take the small $|\vec x|$ expansion (\ref{smallx}). We find that, to lowest order in $x^i$, the solution is
\begin{align}
	a_- = \frac{1}{24}\Omega_{ik}\eta^a_{kj}\sigma^a x^ix^j\left(\mu(x^-)\partial_-^2 \mu(x^-) - (\partial_- \mu(x^-))^2\right)+\ldots \ , 
\end{align}
where the ellipsis denotes higher order powers of $x^i$. We assume that there are solutions which remain suitably bounded at $|\vec x|\to\infty$.
In particular $a_-$ is finite and does not affect the  singular nature of the $A_i$ gauge field at $|\vec x|\to 0$. 
For a moving instanton we expect a solution similar to that found in \cite{Lambert:2011gb,Mouland:2019zjr} which behaves as 
$\omega \sim \eta^a_{ij}x^i\partial_-y^j/|\vec x|^2$ and leads to a finite contribution to $A_i$. We also note that  $a_-=0$ at transition points where $\mu=\dot\mu=0$.

  The first term in the action can then in principle be evaluated to give an expression involving $\Phi$ and $\partial_-\Phi$ given in terms of multiple integrals of $\mu_A(\tau)$   over the instanton  worldlines. We leave this as an exercise to the enthusiastic reader.
 
Next we look at the remaining terms in the action which are all boundary terms arising from scalar fields which are given by (\ref{scalars}). For simplicity we set $A_{0-}=0$.  For a  generic scalar solution of the Laplacian, which we simply denote by $X$, in a spin $s$ representation of the  $SU(2)$ gauge group, we have
\begin{align}\label{expansion}
X^\dag \hat D_i X = -sX_{0}^\dag X_{0} \frac{\hat \partial_i\Phi}{\Phi^{2s+1}}	 + i\eta^a_{ik}X_{0}^\dag T_a(X_{0})\frac{\hat \partial_k\Phi}{\Phi^{2s+1}}\ .
\end{align}

First we consider the component of the boundary at $|\vec x|\to\infty$. As $|\vec x|\to \infty$   we found above that
\begin{align}
	\Phi \to  1 + \frac{2\pi R}{|\vec x|^2}\sum_A(\mu_A(\infty)+\mu_A(-\infty))+\ldots\ ,
\end{align}
where the ellipsis refers to lower order terms in $1/|\vec x|$. 
The first term in (\ref{expansion}) leads to a contribution 
\begin{align}\label{Vis}
 -sX_{0}^\dag X_{0}\int_{|\vec x|\to\infty} dx^-d\Omega_3  	|\vec x|^2 x^i \frac{\hat \partial_i\Phi}{\Phi^{2s+1}}& =  -sX_{0}^\dag X_{0}\int_{|\vec x|\to\infty} dx^-d\Omega_3  	|\vec x|^2 \frac{ x^i\partial_i\Phi}{\Phi^{2s+1}}\nonumber\\ 
 &=  {8\pi^3 R s} X_{0}^\dag X_{0} \int dx^- 	\sum_A ( \mu_A(\infty)+\mu_A(-\infty)) \ .
\end{align}
Thus to obtain a finite action we require $X_0=0$ or $\mu_A(\pm\infty)=0$. This latter condition can be thought of as the requirement that there are no instantons present at $x^-\to\pm\infty$ (although there can be solutions where there are instantons at any finite value of $x^-$, just with a size that shrinks to zero as in (\ref{ex2})).
 
 From the second term in  (\ref{expansion}) we find
 \begin{align} 	
 i\eta^a_{ik}X_{0}^\dag T_a(X_{0})\int_{|\vec x|\to\infty} dx^-d\Omega_3  	|\vec x|^2 x^i \frac{\hat \partial_k\Phi}{\Phi^{2s+1}}	 &=i\eta^a_{ik}X_{0}^\dag T_a(X_{0})\int_{|\vec x|\to\infty} dx^-d\Omega_3  	|\vec x|^2  \frac{x^i }{\Phi^{2s+1}}\nonumber\\ &\hskip4cm\times \left(\partial_k \Phi - \frac12 \Omega_{kj}x^j\partial_-\Phi\right) \nonumber\\
 &=-\frac{i}{2}\eta^a_{ik}\Omega_{kj}X_{0}^\dag T_a(X_{0})\int_{|\vec x|\to\infty} dx^-d\Omega_3  	|\vec x|^2 x^i x^j \frac{\partial_-\Phi }{\Phi^{2s+1}} \ ,
  \end{align}
where the $\partial_k\Phi$ contribution vanishes since $\partial_k\Phi \sim x_k/|\vec x|^4+\ldots$. 
Thus so long as $\partial_-\Phi\to 0$ faster than $1/|\vec x|^{4}$ this contribution will vanish. 

 Next we consider 
the contributions from the $x^-\to\pm\infty$ boundary pieces. Note that $0\le \Phi^{-s}\le 1$ so the only divergences in the integrals arise from the $|\vec x|\to\infty$ region.  To begin with we  have contributions from the first term in (\ref{expansion}):
\begin{align}
\frac{s}2 X^\dag_0X_0\Omega_{ij}\int_{x^-\to \pm\infty} d^4 x x^j	\frac{\hat \partial_i\Phi}{\Phi^{2s+1}} &= \frac{s}2 X^\dag_0X_0\Omega_{ij}\int_{x^-\to \pm\infty} d^4 x x^j	\frac{\partial_i\Phi}{\Phi^{2s+1}}\nonumber\\
&\hskip 2cm - \frac{s}{4R^2} X^\dag_0X_0 \int_{x^-\to\pm\infty} d^4 x |\vec x|^2 	\frac{\partial_-\Phi}{\Phi^{2s+1}}\ .
\end{align}
For spherically symmetric solutions the first term vanishes. More generally  we find, as $x^-\to\pm\infty$,
\begin{align}\label{LPhi}
	\Phi \to  1 + \sum_A\frac{4\pi R\mu_A(\pm\infty)}{|\vec x-\vec y_A(\pm\infty)|^2} + \ldots\ ,
	\end{align}
where the ellipsis denotes lower order terms in $1/|\vec x|$. So the 
	first term is  convergent if $\mu_A(\pm\infty)=0$.
	 The second term will be convergent if  $\partial_-\Phi\to 0$ faster than $1/|\vec x|^{6}$. 
	 
	 Lastly we  have the contributions from the second term in (\ref{expansion}):
\begin{align}
\frac{i}{2} X^\dag_0T_a(X_0)\Omega_{ij}\int_{x^-\to \pm\infty} d^4 x x^j\eta_{ik}^a	\frac{\hat \partial_k\Phi}{\Phi^{2s+1}} &= \frac{i}{2} X^\dag_0T_a(X_0)\Omega_{ij}\eta_{ik}\int_{x^-\to \pm\infty} d^4 x x^j	\frac{\partial_k\Phi}{\Phi^{2s+1}}\ .
\end{align}
Again given the form  (\ref{LPhi}) with  $\mu_A(\pm\infty)=0$ at leading order we encounter  integrals  of the form
\begin{align}\label{symint}
	\int d^4x x^jx^k F(|\vec x|^2) = \frac14\delta^{jk} \int d^4x |\vec x|^2F(|\vec x|^2)\ ,\end{align}
for a suitable choice of $F$,  whose contribution will therefore vanish as $\eta^a_{ij}\Omega_{ij}=0$.

Thus in summary,  if $\mu_A(\pm\infty)=0$ and $\partial_-\Phi\to 0$ faster than $1/|\vec x|^6$ then the scalar field contributions to the action from the $|\vec x|\to\infty$ boundary component vanish and the contributions from the $x^-\to\pm \infty$ boundary components are finite. 
In particular these conditions are satisfied by the solution (\ref{ex2}) as $\mu(\pm\infty)=0$  and $\partial_-\Phi \sim 1/|\vec x|^8$ as $|\vec x|\to\infty$.

\subsection{A Curious Exact Solution}

Remarkably, assuming the 't Hooft ansatz, we can find an exact form for $A_-$ that solves (\ref{A-eq}):
 \begin{align}\label{omegasol}
 A_- &= -\frac{1}{4}R^2\Omega_{ij}{\hat F}_{ij}- \frac{i}{2}R^2\Omega_{ij}[\hat A_i,\hat A_j] 
 \nonumber\\
 & = \frac{1}{4}R^2\Omega_{ik}\eta^a_{kj}\sigma^a\Phi^{-1}\hat \partial_i\hat\partial_j \Phi	 \ .
 \end{align}
 In addition we have the option to add zero-modes such as $u\Phi^2+v\Phi^{-1}$
 where $u,v$ are  constant $su(2)$ matrices. However a non-zero $u$ leads to singular configurations whereas  solutions  with $v$ non-zero do not  change our discussion below.  This solution is notable as it means that we have explicitly solved all the dynamical field equations in terms of the function $\Phi$.  It would be interesting to know if a similar solution exists more generally, beyond the 't Hooft ansatz.
 
 Furthermore we find that near a worldline, which we take to be at $x^i=0$,
 \begin{align}
 	A_i = \eta^a_{ij}\frac{x^j}{|\vec x|^2} - R^2\Omega_{il}\eta^a_{km}\Omega_{kn}\frac{x^lx^mx^n}{|\vec x|^4}+\ldots
 \end{align} 
 The extra contribution to the singularity in the gauge field actually cancels the instanton number arising from the first term! More precisely, we find that as we approach $|\vec x|\to 0$, the Chern-Simons 3-form goes as $\nu_3|_{S^3_0}=\mathcal{O}(|\vec x|^{-2})d\Omega_3$, in contrast to the finite behaviour found previously. Noting further that this solution for $A_-$ dies away as $|\vec x|\to\infty$ sufficiently fast to not affect the contribution to $Q$ from the integral at $S^3_\infty$, we find $Q=0$. Note that the second term on its own does not define a gauge field with instanton number, but adding it to the anti-instanton removes the instanton.  Thus we find exact solutions given by $\Phi$ but all with vanishing instanton number for the original gauge field strength $F_{ij}$. These solutions presumably still can be interpreted as some sort of worldline as the energy density is peaked along a curve $(x^-(\tau),x^i(\tau))$.
 
 Note that in this case $A_-$ does not vanish if $\partial_-\hat A_i =0$. As such it doesn't represent a solution for $\omega$ that was introduced in (\ref{A'-}). Rather it must be identified with $\omega+ A'_-$ for the  $\omega$  as defined and some $A_-'$. On the other hand we argued above that we also expect there to exist classical solutions where $\omega$ does not affect the gauge field instanton number. For example if we consider the static case then we see that there are at least two acceptable solutions for $A_-$ (and again we could include the zero-modes). One is simply $A_-=0$ in which case $A_i=\hat A_i$ and we indeed find the $A_i$ has a non vanishing instanton number. However we can also take $A_-$ to be given by (\ref{omegasol}) in which case $A_i$ does not carry any instanton number, although $\hat A_i$ remains the same. For non-static solutions  (\ref{omegasol}) is a valid solution again leading to a vanishing instanton number for the gauge field $A_i$. However we have argued that in this case there also exists a solution for $\omega$ such that 
 (\ref{A-eq}) is solved and the instanton number of $A_i$ is non-vanishing.

\subsection{Recovering $\Omega_{ij}=0$}\label{subsec: DLCQ}

 Lastly let us consider the  $\Omega_{ij}=0$ case, corresponding to $R\to\infty$,  which was studied in \cite{Lambert:2011gb,Mouland:2019zjr}. Here the  constraint simply states that $A_i$ has an anti-self-dual field strength on ${\mathbb R}^4$. As such it is determined in complete generality by the ADHM construction as an explicit function of $\vec x$ as well as a finite set of moduli $m^I$. As far as the constraint is concerned these moduli can depend arbitrarily on $x^-$. The action and equations of motion take the same form but now $\hat A_i=A_i$ and $\hat D_i=D_i$. We can then write
 \begin{align}
 \partial_- A_i = \partial_I A_i \partial_- m^I	\ ,
 \end{align}
and we expand
\begin{align}
	a_- = \omega_I\partial_-m^I\ ,
\end{align}
so
that (\ref{A'-}) becomes 
\begin{align}
 D_iD_i \omega_I = D_i\partial_I A_i\ .
\end{align}
The interpretation is that  $a_-$ acts as a compensating gauge transformation which ensures that
\begin{align}
\delta A_i =( \partial_IA_i - D_i\omega_I)\delta m^I	 \ , \qquad \delta m^I = \partial_-m^I\delta x^-\ ,
\end{align}
is orthogonal to a gauge transformation in the sense that:
\begin{align}
D_i\delta A_i =0	\ .
\end{align}
 In this way $\delta A_i$ can be viewed as a tangent vector to the moduli space of anti-self-dual gauge fields (see \cite{Tong:2005un}).  We are not aware of any closed form expression for $a_-$ in the $\Omega_{ij}=0$ case\footnote{The solution (\ref{omegasol}) diverges in the $\Omega_{ij}\to0$ limit.}. 
 
 If we now evaluate the action (\ref{S}) we find (still assuming $A'_-=0$)
 \begin{align}
S = \frac{1}{g^2_{YM}}\int dx^- \Big[  \frac12 g_{IJ}\partial_- m^I \partial_- m^J - V\Big] \ ,	
\end{align}
where the moduli space metric is defined by 
\begin{align}\label{metric}
	  g_{IJ} = \int d^4x\ {\rm tr} \left((  \partial_I A_i - D_i\omega_I)(  \partial_J A_i - D_i\omega_J)\right)\ .
\end{align} 
and now we find a potential for the moduli $V$ that comes from the scalar field boundary terms (\ref{Vis})
\begin{align}
V &= \sum_{scalars}\int d\Omega_3 |\vec x| x^i X^\dag D_i X	\nonumber\\
& = 4\pi^2 \left(\sum_{scalars} s_X X^\dag_0X_0\right) \sum_A \rho^2_A\ ,
\end{align}
where in the second line we evaluated the integral using the standard 't Hooft ansatz obtained by taking linear combinations of the solution (\ref{S'tH}) but translated to have poles at points $\vec y_A \in {\mathbb R}^4$. In this case only the boundary components at $|\vec x|\to \infty$ arise. Furthermore the $\rho_A$ are the size moduli. These, along with  $\vec y_A$ and the gauge embedding moduli, are allowed to be arbitrary functions of $x^-$.   Thus we recover the  conventional description of dynamics on the moduli instanton space. More generally one finds that allowing for a non-vanishing $A'_-$ leads to a connection on the moduli space \cite{Lambert:2011gb}.  

Thus for  $\Omega_{ij}= 0$  the dynamics takes place on the moduli space of anti-self-dual gauge fields, as proposed in \cite{Aharony:1997th,Aharony:1997pm}. This space is a disconnected sum with each component labelled by  instanton number and parameterised by a set of positions, sizes and gauge embedding moduli, which we have denoted by $m^I$, all of which are dynamical. Finite action configurations consist of fluctuations of all the moduli in each connected component subjected to a potential for their size when the scalars have a vacuum expectation value.

We know that in the $R\to\infty$ limit, the constraint equation reduces to the usual spatial anti-instanton equation $F_{ij}=-\left(\star F\right)_{ij}$. It is worth therefore asking which solutions to these equations are found in the $R\to\infty$ limit of our 't Hooft solutions at finite $R$.

So consider the $N$ worldline solution (\ref{eq: Phi general solution}). As discussed in Section \ref{sec: gauge field topology}, we can without loss of generality take each of these worldlines to be monotonic and disjoint except for possibly at their endpoints. Then let us consider the $R\to\infty$ limit. Much of our work is already done, since the $R\to\infty$ limit is very similar to the limit in which we approach the worldline, $|\vec{x}-\vec{y}|\to 0$, where the integrals over the $\tau_A$ localise. In particular, one can take the limit explicitly by making use of the Fourier techniques in Section \ref{sec: gauge field topology}. Then, noting that in the limit $\hat{A}_i = A_i$ and $\hat{\partial}_i=\partial_i$, and further normalising as $\mu_A(\tau_A)=(\rho_A(\tau-A))^2/4\pi R$ for some functions $\rho_A(\tau)$, we find that as $R\to\infty$,
\begin{align}
  A_i = \hat{A}_i = -\frac{1}{2}\eta^a_{ij} \sigma^a \partial_j \log \Phi,\qquad \Phi(x) = 1+\sum_{A=1}^N \frac{(\rho_A(x^-))^2}{|\vec x - \vec y_A(x^-)|^2} \ .
  \label{eq: DLCQ limit of t Hooft solution}
\end{align}
Thus, we precisely recover the usual 't Hooft ansatz, in which the size modulus $\rho(x^-)$ and position moduli $y^i(x^-)$ are allowed to vary arbitrarily with time. The solution therefore describes a set of $N$ anti-instantons moving arbitrarily. Interestingly, we can still take any of the $\rho_A$ to have compact support, resulting in anti-instantons that are created and annihilated. Indeed, all of the analysis of Section \ref{sec: gauge field topology} persists in the $R\to\infty$ limit, and is indeed much more immediate due to local form of (\ref{eq: DLCQ limit of t Hooft solution}).

It is interesting that at finite $R$, the value of the gauge field at some point $x=(x^-, \vec{x})$ away from worldlines is determined by the moduli $\mu_A(\tau), \vec{y}_A(\tau)$ at \textit{every} point along every worldline, while in the $R\to \infty$ limit the solution localises, in the sense that the gauge field now depends only on the $\rho_A(\tau),\vec{y}_A(\tau)$ at $\tau=x^-$.

Finally, let us consider the fate of our solution (\ref{omegasol}) for $A_-$ in the $R\to\infty$ ($\Omega_{ij}\to 0$) limit. In fact, the solution (\ref{omegasol}) diverges, however $R^{-1}A_-$ is finite and becomes harmonic: $\partial_i\partial_i (R^{-1}A_-)\to 0$. Therefore  the combination
\begin{align}
A_i &= \hat A_i + \frac12 \Omega_{ij}x^j A_- \nonumber\\
&=  	 \hat A_i + \frac12 R\Omega_{ij}x^j \frac{1}{R}A_-\ , 
\end{align}
will remain finite and in the limit becomes 
\begin{align}
A_i	= -\frac{1}{2\Phi}\eta^a_{ij}\sigma^a  \partial_j \Phi+ \frac{1}{8\Phi}\Omega'_{ij}x^j \Omega'_{mk}\eta^a_{kn}\sigma^a  \partial_m \partial_n \Phi	\ ,
\end{align}
where $\Omega' = R\Omega_{ij}$ is a constant non-degenerate anti-self-dual tensor on ${\mathbb R}^4$ and $\Phi$ is harmonic: $\partial_i\partial_i\Phi=0$.  The first term is the usual 't Hooft ansatz solution and carries an instanton number of minus one from each of the poles in $\Phi$.	As we have seen the two terms together give a  solution which does not carry any instanton number. This suggests that even in the ordinary instanton case there is an $\Omega'$-deformation which removes the instanton singularities   by shifting $A_i$
 \begin{align}
 A_i\to A_i +\frac12 \Omega'_{ij}x^jA_-''	\ ,
 \end{align}
 where 
 \begin{align}
A_-'' = \frac{1}{4\Phi} \Omega'_{mk}\eta^a_{kn}\sigma^a  \partial_m \partial_n \Phi\ ,	
\end{align}
 is a harmonic function which is not spherically symmetric, even if $\Phi$ is.
 The only difference is that there is no preferred choice for $ \Omega'_{ij}$ but rather a three-dimensional family of choices.

  \section{Discussion and Conclusion}
  
  In this paper we discussed the dynamics of the Bosonic sector of a class of supersymmetric five-dimensional non-abelian gauge theories without Lorentz symmetry but which admit an $SU(1,3)$ conformal symmetry. In particular we showed that a generalised 't Hooft ansatz linearised the anti-self-duality constraint and allowed us to construct a wide class of solutions. These have a physical interpretation as representing instanton worldlines where the position and size are allowed to evolve. In particular we presented finite action examples of configurations where the instantons shrink to zero size. We interpreted these as the creation and an annihilation of instantons in the gauge theory. 
  
  In addition to studying the constraint surface we also examined solutions to the equations of motion on the constraint surface. This in turn involved solving  for $a_-$ in (\ref{A'-}). A general closed form expression for $a_-$ seems out of reach however we argued that there are solutions to the equations of motion which do not change the nature of the original gauge field near the worldline. In addition we also found a class of exact solutions given by (\ref{omegasol}) which modify the singularity of the original gauge field $A_i$ leading to configurations with zero instanton number.
  We also showed that the scalar fields can be solved for in the 't Hooft ansatz and discussed their contribution to the action. 
  
  For $\Omega_{ij}=0$ we recover the familiar results where the constraint surface consists of anti-self-dual gauge fields and the dynamics reduces to motion on instanton moduli space. On the other hand for $\Omega_{ij}\ne 0$ our 't Hooft ansatz presents a picture where the constraint surface consists of dynamical instanton solutions where the size and positions of the instantons evolve along a worldline. The worldlines can take any reasonable form and the self-duality condition computes the backreaction. Presumably in a more general solution the gauge embedding moduli will also be allowed to fluctuate along the worldline. In this case  we found  that a finite action requires that the instantons all shrink to zero size at $x^-\to \pm\infty$. Thus it seems as if the action for  $\Omega_{ij}\ne 0$ describes the dynamics of the creation and annihilation of instantons, whereas at $\Omega_{ij}=0$, once we reduce the action to motion on the moduli space, the dynamics takes place on a background of fixed instanton number. In the usual DLCQ prescription the full theory is  captured by considering the various instanton sectors separately, although in this case too one could introduce instanton operators.  It is also worth noting that for $\Omega_{ij}=0$ the  scalars contributed to the action {\it via} a potential arising from a boundary term located at $|\vec x|\to \infty$. Whereas for $\Omega_{ij}\ne0$ such boundary contributions lead to a divergent action unless we took the instantons to vanish at $x^-\to\pm\infty$. Instead we found that the contribution of the scalars to the action comes entirely from the boundaries at $x^-\to\pm\infty$.
  
  It would be interesting to obtain a complete description of the solutions to the anti-self-dual gauge field constraint, generalising the familiar ADHM construction, and to explore a holographic interpretation in the large-rank limit analogous to the analysis for $\mathcal{N}=4$ super-Yang-Mills \cite{Dorey:1999pd}. We also need to include the fermions and incorporate the effects of supersymmetry. In future work  \cite{US} we will report on the properties of instanton operators in the quantum theory that can create and annihilate instantons and which play a central role in extending the symmetry, following the analysis of \cite{Lambert:2014jna,Tachikawa:2015mha,Bergman:2016avc}. It would very interesting if this could provide a systematic way to compute correlators in the $\Omega$-deformed five-dimensional theory which goes beyond the constraints of superconformal symmetry and does not rely on dimensional reduction. Indeed, one would hope to go in the reverse direction by computing general correlation functions in the five-dimensional theory using path integral methods and then assembling them into a Fourier series corresponding to six-dimensional correlators. Lastly the large amounts of symmetry along with the linearisation of the anti-self-duality constraint suggests that there maybe a role for integrability techniques. 
  
 \section*{Aknowledgements} 
 
P.~Richmond was supported   by STFC grant ST/L000326/1, A.~Lipstein by the Royal Society as a Royal Society University Research Fellowship holder and  R.~Mouland by the STFC studentship ST10837.

\bibliographystyle{JHEP}
\bibliography{OmInst}

\providecommand{\href}[2]{#2}\begingroup\raggedright\begin{thebibliography}{10}

\bibitem{Nahm:1977tg}
W.~Nahm, \emph{{Supersymmetries and their Representations}},
  \href{https://doi.org/10.1016/0550-3213(78)90218-3}{\emph{Nucl. Phys. B}
  {\bfseries 135} (1978) 149}.

\bibitem{Witten:1995zh}
E.~Witten, \emph{{Some comments on string dynamics}},  in \emph{{STRINGS 95:
  Future Perspectives in String Theory}}, pp.~501--523, 7, 1995,
  \href{https://arxiv.org/abs/hep-th/9507121}{{\ttfamily hep-th/9507121}}.

\bibitem{Strominger:1995ac}
A.~Strominger, \emph{{Open p-branes}},
  \href{https://doi.org/10.1016/0370-2693(96)00712-5}{\emph{Phys. Lett. B}
  {\bfseries 383} (1996) 44}
  [\href{https://arxiv.org/abs/hep-th/9512059}{{\ttfamily hep-th/9512059}}].

\bibitem{Seiberg:1996qx}
N.~Seiberg, \emph{{Nontrivial fixed points of the renormalization group in
  six-dimensions}},
  \href{https://doi.org/10.1016/S0370-2693(96)01424-4}{\emph{Phys. Lett. B}
  {\bfseries 390} (1997) 169}
  [\href{https://arxiv.org/abs/hep-th/9609161}{{\ttfamily hep-th/9609161}}].

\bibitem{Lambert:2019jwi}
N.~Lambert, A.~Lipstein and P.~Richmond, \emph{{Non-Lorentzian M5-brane
  Theories from Holography}},
  \href{https://doi.org/10.1007/JHEP08(2019)060}{\emph{JHEP} {\bfseries 08}
  (2019) 060} [\href{https://arxiv.org/abs/1904.07547}{{\ttfamily
  1904.07547}}].

\bibitem{Lambert:2020jjm}
N.~Lambert and T.~Orchard, \emph{{Non-Lorentzian Avatars of (1,0) Theories}},
  \href{https://arxiv.org/abs/2011.06968}{{\ttfamily 2011.06968}}.

\bibitem{Aharony:1997th}
O.~Aharony, M.~Berkooz, S.~Kachru, N.~Seiberg and E.~Silverstein, \emph{{Matrix
  description of interacting theories in six-dimensions}},
  \href{https://doi.org/10.4310/ATMP.1997.v1.n1.a5}{\emph{Adv. Theor. Math.
  Phys.} {\bfseries 1} (1998) 148}
  [\href{https://arxiv.org/abs/hep-th/9707079}{{\ttfamily hep-th/9707079}}].

\bibitem{Aharony:1997pm}
O.~Aharony, M.~Berkooz, S.~Kachru and E.~Silverstein, \emph{{Matrix description
  of (1,0) theories in six-dimensions}},
  \href{https://doi.org/10.1016/S0370-2693(97)01503-7}{\emph{Phys. Lett. B}
  {\bfseries 420} (1998) 55}
  [\href{https://arxiv.org/abs/hep-th/9709118}{{\ttfamily hep-th/9709118}}].

\bibitem{Lambert:2020zdc}
N.~Lambert, A.~Lipstein, R.~Mouland and P.~Richmond, \emph{{Five-Dimensional
  Non-Lorentzian Conformal Field Theories and their Relation to
  Six-Dimensions}}, \href{https://doi.org/10.1007/JHEP03(2021)053}{\emph{JHEP}
  {\bfseries 03} (2021) 053}
  [\href{https://arxiv.org/abs/2012.00626}{{\ttfamily 2012.00626}}].

\bibitem{Lambert:2019fne}
N.~Lambert, A.~Lipstein, R.~Mouland and P.~Richmond, \emph{{Bosonic symmetries
  of $(2,0)$ DLCQ field theories}},
  \href{https://doi.org/10.1007/JHEP01(2020)166}{\emph{JHEP} {\bfseries 01}
  (2020) 166} [\href{https://arxiv.org/abs/1912.02638}{{\ttfamily
  1912.02638}}].

\bibitem{Vandoren:2008xg}
S.~Vandoren and P.~van Nieuwenhuizen, \emph{{Lectures on instantons}},
  \href{https://arxiv.org/abs/0802.1862}{{\ttfamily 0802.1862}}.

\bibitem{Aronszajn1956}
N.~Aronszajn and W.~F. Donoghue, \emph{{On exponential representations of
  analytic functions in the upper half-plane with positive imaginary part}},
  \href{https://doi.org/10.1007/BF02937349}{\emph{Journal d'Analyse
  Mathematique} {\bfseries 5} (1956) 321}.

\bibitem{Lambert:2014jna}
N.~Lambert, C.~Papageorgakis and M.~Schmidt-Sommerfeld, \emph{{Instanton
  Operators in Five-Dimensional Gauge Theories}},
  \href{https://doi.org/10.1007/JHEP03(2015)019}{\emph{JHEP} {\bfseries 03}
  (2015) 019} [\href{https://arxiv.org/abs/1412.2789}{{\ttfamily 1412.2789}}].

\bibitem{Tachikawa:2015mha}
Y.~Tachikawa, \emph{{Instanton operators and symmetry enhancement in 5d
  supersymmetric gauge theories}},
  \href{https://doi.org/10.1093/ptep/ptv040}{\emph{PTEP} {\bfseries 2015}
  (2015) 043B06} [\href{https://arxiv.org/abs/1501.01031}{{\ttfamily
  1501.01031}}].

\bibitem{Bergman:2016avc}
O.~Bergman and D.~Rodriguez-Gomez, \emph{{A Note on Instanton Operators,
  Instanton Particles, and Supersymmetry}},
  \href{https://doi.org/10.1007/JHEP05(2016)068}{\emph{JHEP} {\bfseries 05}
  (2016) 068} [\href{https://arxiv.org/abs/1601.00752}{{\ttfamily
  1601.00752}}].

\bibitem{US}
N.~Lambert, A.~Lipstein, R.~Mouland and P.~Richmond, \emph{{to appear}}, .

\bibitem{Arutyunov:2002ff}
G.~Arutyunov and E.~Sokatchev, \emph{{Implications of superconformal symmetry
  for interacting (2,0) tensor multiplets}},
  \href{https://doi.org/10.1016/S0550-3213(02)00359-0}{\emph{Nucl. Phys. B}
  {\bfseries 635} (2002) 3}
  [\href{https://arxiv.org/abs/hep-th/0201145}{{\ttfamily hep-th/0201145}}].

\bibitem{Heslop:2004du}
P.~J. Heslop, \emph{{Aspects of superconformal field theories in six
  dimensions}},
  \href{https://doi.org/10.1088/1126-6708/2004/07/056}{\emph{JHEP} {\bfseries
  07} (2004) 056} [\href{https://arxiv.org/abs/hep-th/0405245}{{\ttfamily
  hep-th/0405245}}].

\bibitem{Rastelli:2017ymc}
L.~Rastelli and X.~Zhou, \emph{{Holographic Four-Point Functions in the (2, 0)
  Theory}}, \href{https://doi.org/10.1007/JHEP06(2018)087}{\emph{JHEP}
  {\bfseries 06} (2018) 087}
  [\href{https://arxiv.org/abs/1712.02788}{{\ttfamily 1712.02788}}].

\bibitem{Heslop:2017sco}
P.~Heslop and A.~E. Lipstein, \emph{{M-theory Beyond The Supergravity
  Approximation}}, \href{https://doi.org/10.1007/JHEP02(2018)004}{\emph{JHEP}
  {\bfseries 02} (2018) 004}
  [\href{https://arxiv.org/abs/1712.08570}{{\ttfamily 1712.08570}}].

\bibitem{Chester:2018dga}
S.~M. Chester and E.~Perlmutter, \emph{{M-Theory Reconstruction from (2,0) CFT
  and the Chiral Algebra Conjecture}},
  \href{https://doi.org/10.1007/JHEP08(2018)116}{\emph{JHEP} {\bfseries 08}
  (2018) 116} [\href{https://arxiv.org/abs/1805.00892}{{\ttfamily
  1805.00892}}].

\bibitem{Abl:2019jhh}
T.~Abl, P.~Heslop and A.~E. Lipstein, \emph{{Recursion relations for anomalous
  dimensions in the 6d $(2, 0)$ theory}},
  \href{https://doi.org/10.1007/JHEP04(2019)038}{\emph{JHEP} {\bfseries 04}
  (2019) 038} [\href{https://arxiv.org/abs/1902.00463}{{\ttfamily
  1902.00463}}].

\bibitem{Alday:2020lbp}
L.~F. Alday and X.~Zhou, \emph{{All Tree-Level Correlators for M-theory on
  $AdS_7 \times S^4$}},
  \href{https://doi.org/10.1103/PhysRevLett.125.131604}{\emph{Phys. Rev. Lett.}
  {\bfseries 125} (2020) 131604}
  [\href{https://arxiv.org/abs/2006.06653}{{\ttfamily 2006.06653}}].

\bibitem{Alday:2020tgi}
L.~F. Alday, S.~M. Chester and H.~Raj, \emph{{6d (2,0) and M-theory at
  1-loop}}, \href{https://doi.org/10.1007/JHEP01(2021)133}{\emph{JHEP}
  {\bfseries 01} (2021) 133}
  [\href{https://arxiv.org/abs/2005.07175}{{\ttfamily 2005.07175}}].

\bibitem{Lambert:2011gb}
N.~Lambert and P.~Richmond, \emph{{(2,0) Supersymmetry and the Light-Cone
  Description of M5-branes}},
  \href{https://doi.org/10.1007/JHEP02(2012)013}{\emph{JHEP} {\bfseries 02}
  (2012) 013} [\href{https://arxiv.org/abs/1109.6454}{{\ttfamily 1109.6454}}].

\bibitem{Mouland:2019zjr}
R.~Mouland, \emph{{Supersymmetric soliton $\sigma$-models from non-Lorentzian
  field theories}}, \href{https://doi.org/10.1007/JHEP04(2020)129}{\emph{JHEP}
  {\bfseries 04} (2020) 129}
  [\href{https://arxiv.org/abs/1911.11504}{{\ttfamily 1911.11504}}].

\bibitem{Tong:2005un}
D.~Tong, \emph{{TASI lectures on solitons: Instantons, monopoles, vortices and
  kinks}},  in \emph{{Theoretical Advanced Study Institute in Elementary
  Particle Physics}: {Many Dimensions of String Theory}}, 6, 2005,
  \href{https://arxiv.org/abs/hep-th/0509216}{{\ttfamily hep-th/0509216}}.

\bibitem{Dorey:1999pd}
N.~Dorey, T.~J. Hollowood, V.~V. Khoze, M.~P. Mattis and S.~Vandoren,
  \emph{{Multi-instanton calculus and the AdS / CFT correspondence in N=4
  superconformal field theory}},
  \href{https://doi.org/10.1016/S0550-3213(99)00193-5}{\emph{Nucl. Phys. B}
  {\bfseries 552} (1999) 88}
  [\href{https://arxiv.org/abs/hep-th/9901128}{{\ttfamily hep-th/9901128}}].

\end{thebibliography}\endgroup

\end{document}